%% file: neurips_2020.tex
\documentclass{article}

\usepackage[preprint, nonatbib]{neurips_2020}

\usepackage[utf8]{inputenc} % allow utf-8 input
\usepackage[T1]{fontenc}    % use 8-bit T1 fonts
\usepackage{hyperref}       % hyperlinks
\usepackage{url}            % simple URL typesetting
\usepackage{booktabs}       % professional-quality tables
\usepackage{amsfonts}       % blackboard math symbols
\usepackage{nicefrac}       % compact symbols for 1/2, etc.
\usepackage{microtype}      % microtypography
\usepackage{amsmath} 
\usepackage[square,numbers]{natbib}
\usepackage{amsthm}
\usepackage{xcolor}

\usepackage{graphicx}
\usepackage{tikz}
\usetikzlibrary{decorations.pathreplacing}
\usepackage[utf8]{inputenc}
\usepackage{pgfplots}
\usepackage{xcolor}
\usepackage{subcaption} %  for subfigures environments 

\theoremstyle{remark}

\usepackage{algorithm,algpseudocode}% http://ctan.org/pkg/{algorithms,algorithmicx}

\title{Meta Learning MPC using Finite-Dimensional Gaussian Process Approximations }

\author{%
  Elena Arcari  \\
  Institute for Dynamic Systems and Control\\
  ETH Zurich\\
  \texttt{earcari@ethz.ch} \\
  \And
  Andrea Carron \\
  Institute for Dynamic Systems and Control\\
  ETH Zurich\\
  \texttt{carrona@ethz.ch} \\
  \AND
 Melanie N. Zeilinger \\
 Institute for Dynamic Systems and Control\\
  ETH Zurich\\
 \texttt{mzeilinger@ethz.ch} \\
}
{}
\begin{document}

\begin{center}
{\huge{Cover Note}}
\end{center}

\begin{center}
{\Large{\today}}
\end{center}

{\Large{
The results presented in this paper are preliminary. \\
A new version that discusses an updated methodology, and shows new simulation and hardware results on a different robotic platform, can be found here: \\
\url{https://arxiv.org/abs/2211.10270}}}

\newpage

\maketitle

\begin{abstract}
  Data availability has dramatically increased in recent years, driving model-based control methods to exploit learning techniques for improving the system description, and thus control performance. Two key factors that hinder the practical applicability of learning methods in control are their high computational complexity and limited generalization capabilities to unseen conditions. Meta-learning is a powerful tool that enables efficient learning across a finite set of related tasks, easing adaptation to new unseen tasks. This paper makes use of a meta-learning approach for adaptive model predictive control, by learning a system model that leverages data from previous related tasks, while enabling fast fine-tuning to the current task during closed-loop operation. The dynamics is modeled via Gaussian process regression and, building on the Karhunen-Lo{\`e}ve expansion, can be approximately reformulated as a finite linear combination of kernel eigenfunctions. Using data collected over a set of tasks, the eigenfunction hyperparameters are optimized in a meta-training phase by maximizing a variational bound for the log-marginal likelihood. During meta-testing, the eigenfunctions are fixed, so that only the linear parameters are adapted to the new unseen task in an online adaptive fashion via Bayesian linear regression, providing a simple and efficient inference scheme. Simulation results are provided for autonomous racing with miniature race cars adapting to unseen road conditions.
\end{abstract}

\section{Introduction}

Data-driven model-based control techniques have gained increasing attention, most commonly focusing on improving system knowledge during operation. While this enhances control performance, data-driven control methods come with important challenges~\cite{LH19b}, particularly regarding the trade-off between model complexity and computational requirements. Gaussian process (GP) regression, e.g., offers appealing properties for modeling complex interactions, but scales poorly, and can be potentially intractable for real-time operation. For this reason, existing work~\cite{JK19}, \cite{YP17}, \cite{CJO14}, \cite{SK18}, \cite{MD13}, \cite{CDK19} reverted to the use of sparse GP approaches~\cite{MT09},\cite{QC05} or techniques for approximating GPs along the control horizon~\cite{QC03}. These methods are typically based on collecting data for a single task, where the learning algorithm improves the model based on online data, under the assumption that the task will not change. This approach naturally limits adaptation to changing conditions, for instance in the case of a car driving on dry and wet asphalt, or a robotic manipulator tracking a given reference trajectory that changes from task to task, which is often addressed by means of heuristic data selection procedures. Meta-learning~\cite{CF17} systematically addresses this issue with the goal of transferring acquired knowledge to unseen tasks with minimal new data.

In this work, we exploit the strength of meta-learning for enabling data-efficient adaptive model predictive control{} (MPC), making use of GP regression for modeling the system dynamics. The idea is to leverage the Karhunen-Lo{\`e}ve expansion~\cite{GP18} and approximate the GP as a finite linear combination of basis functions. In a meta-learning fashion, the associated hyperparameters are meta-trained over a set of related tasks making use of a variational bound for the log-marginal likelihood, which is typically used for training GPs~\cite{CR17}. During meta-testing, the basis functions are fixed and only the linear parameters are adapted online to a new task via Bayesian linear regression~\cite{CB06}. We analyse and validate the proposed method in simulation on two different control problems. The first is the mountain-car problem~\cite{RS18}, where the tasks are specified as different slopes. In the second example, we consider the control of a miniature race car along a track under changing road conditions.

The contributions of this paper are twofold:
\begin{itemize}
  \item We develop a meta-learning approach for adaptive MPC, providing both generalization capabilities to unseen conditions, and a low complexity inference scheme where at meta-testing only the linear parameters are adapted via Bayesian linear regression.
  \item We provide a meta-training formulation that extends to eigenfunction approximations whenever these are not readily available in closed-form, such that further hyperparameters can be optimized, e.g. the input locations for a subset of regressors.
\end{itemize}

\section{Related Work}

Meta-learning, also known as \emph{learning to learn}, was first proposed as a model-agnostic procedure~\citep{CF17} that enables to construct a strong prior across tasks, which can then be transferred to new unseen tasks. Given the versatility of the approach, it has naturally been extended, and reformulated, in the context of reinforcement learning (RL), particularly for model-free approaches~\cite{JW16},\cite{ZX19}, \cite{AM18}. These, however, still need large amounts of data to achieve acceptable learning results, which can be impractical for real-world systems, thus motivating a shift towards model-based approaches~\cite{LZ20}. In~\cite{IC19}, the authors propose a meta-learning approach for neural network dynamics models to be adapted online in an RL framework. The main idea is to define a task as a collection of past time-steps, used to meta-learn a model that can be easily adapted to future time-steps. Other model-based approaches were proposed in~\cite{SS18} and~\cite{JH18a}. In~\cite{SS18}, the dynamics are modeled as a latent variable GP, interpreting meta-learning as hierarchical Bayes~\cite{EG18}. While we make use of a similar variational approach for enabling meta-training, our problem is formulated to provide exact probability updates for the task parameters by using Bayesian linear regression, and by employing a linear finite-dimensional GP approximation. The approach developed in~\cite{JH18a} is based on Bayesian last-layer models~\cite{JH18b}, where the features consist in neural networks for which the weights are meta-learned, and only the last layer of linear parameters is adapted to new unseen tasks. This scheme enables efficient online adaptation, similarly to the approach employed in this paper, with the difference that the meta-training phase is formulated as an optimization problem over the posterior predictive distribution for finding a good prior over the linear parameters. In contrast, we optimize over a variational bound of the log-marginal likelihood for learning kernel hyperparameters and, potentially, additional basis function hyperparameters when closed form eigenfunctions are not available.

Another related area of research considers multi-task GPs, e.g.~\cite{KY05}, \cite{XL05}, where in particular the first makes use of finite-dimensional approximations, but, similarly to~\cite{JH18b}, seeks for a good prior over the linear parameters using an expectation-maximization approach. The second is also based on the concept of linear combinations of basis functions, but rather in the context of neural networks and restricted to the case of radial basis functions. An interesting  generalization of the variational inference procedure proposed in this paper, and similarly in~\cite{SS18}, can be found in~\cite{AB19}, where a generic multi-task objective function for deep GPs is formulated as a sum of task specific losses and shared regularizers. In our case, while the employed GP has a shallow structure, the obtained variational bound also contains the terms highlighted in~\cite{AB19}: the task-specific losses are expressed as log-likelihood terms and the shared regularizers are in the form of the KL divergence between the variational distributions associated to each task and the prior over all tasks. In the context of Markov decision processes (MDPs), a related approach is developed in~\cite{FD16}, which is based on~\cite{YT05}, and makes use of GP linear mixture models.

Extensive prior work on online adaptation at test time has been developed in the area of learning-based control and robotics, in particular on the idea of fine-tuning global models with a learned residual~\cite{AC19}, \cite{LH19}, \cite{DTN10} , \cite{RC15} or based on learning local models~\cite{SV00}, \cite{FM14} - which can be interpreted as particular instances of tasks. These methods, however, typically do not use the training time for easing the adaptation at test time as, on the contrary, is done in a meta-learning approach.

\section{Preliminiaries}
\label{sec:preliminaries}

We consider measurement models of the form
\begin{equation*}
  y = f(x) + \epsilon,
\end{equation*}
where $f: \mathcal{X} \rightarrow \mathbb{R}$ is an unknown function mapping from a compact space $\mathcal{X} \in \mathbb{R}^d$, with input locations $x$ distributed as $x \sim \mu(\mathcal{X})$, where $\mu$ is a non-degenerate probability measure on $\mathcal{X}$, and $\epsilon \sim \mathcal{N}(0,\sigma_\epsilon^2)$ is i.i.d. additive measurement noise. The mapping is approximated as a GP, described by a covariance function of the form $k: \mathcal{X} \times \mathcal{X} \rightarrow \mathbb{R}$ as
\begin{equation}
  f \sim GP(0,k(x,x')).
  \label{eq:gp}
\end{equation}
As discussed in~\cite{CR17}, GPs of the form~\eqref{eq:gp} can be expressed as linear combinations of possibly infinite basis functions, and can, therefore, be interpreted as Bayesian linear regression. One possible set of basis functions are the eigenfunctions,
% defined by the following integral equation:
% \begin{equation}
%   \lambda \phi(x) = \int k(x, x') \phi(x) \phi(x') d\mu(x'),
%   \label{eq:integral_equation}
% \end{equation}
% where $\phi(x)$ is an eigenfunction of the kernel $k$ with eigenvalue $\lambda$, with respect to measure $\mu$. Following Mercer's theorem (see e.g.~\cite{HK13}), it is possible to express the kernel in terms of its eigenvalues and eigenfunctions:
% \begin{equation}
%   k(x,x') = \sum_{i=1}^{\infty} \lambda_i \phi_i(x) \phi_i(x'),
%   \label{eq:mercer_thm}
% \end{equation}
% where, assuming a labeling $\phi_1, \phi_2, \dots$ and an ordering $\lambda_1 \geq \lambda_2 \geq \dots >0$, the eigenfunctions are orthogonal with respect to $\mu$, and are chosen to satisfy:
% \begin{equation}
%   \int \phi_i(x) \phi_j(x) d\mu(x) = \delta_{ij},
%   \label{eq:orthogonality}
% \end{equation}
% where $\delta_{ij}$ is the Kronecker delta. Using~\eqref{eq:integral_equation}, \eqref{eq:mercer_thm} and~\eqref{eq:orthogonality} we can reformulate
which allow for reformulating the associated GP $f$ via the Karhunen-Lo{\`e}ve expansion~\cite{GP18}:
\begin{equation}
  f(x) = \sum_{i=1}^{\infty} \phi_i(x) \alpha_i, \quad \alpha_i \sim \mathcal{N}(0,\lambda_i).
  \label{eq:KL_expansion}
\end{equation}
This expansion can be truncated to a finite linear combination of eigenfunctions, by selecting $E$ principal components, or in other words, choosing the best E-dimensional subspace approximation for $f$ in~\eqref{eq:KL_expansion}, see e.g.~\cite{HZ97},
\begin{equation}
  \hat f(x) = \sum_{i=1}^{E}\phi_i(x)\alpha_i, \quad \alpha_i \sim \mathcal{N}(0,\lambda_i).
  \label{eq:E-dim KL}
\end{equation}

Obtaining eigenfunctions %that satisfy the integral equation~\eqref{eq:integral_equation}
under a known input distribution is not an easy task, although there exist important known closed-form solutions, e.g., for a Gaussian input distribution and Gaussian kernel expansion using Hermite polynomials~\cite{GP18}. In practice, it is possible to construct basis functions using, for instance, a subset of regressors or the Nystr{\"o}m method~\cite{CR17}, or random features~\cite{BR08} in the case of stationary kernels.

\section{Method}
\label{sec:method}

Data-efficiency and competitive computation times are key factors for the success of a learning-based controller, and to enable real-time closed-loop operation. The presented approach allows for exploiting data collected from a range of related tasks to quickly adapt online to a new task by learning shared information in a meta-training phase and re-using it during meta-testing. This is achieved by making use of the finite-dimensional GP approximation~\eqref{eq:E-dim KL}, so that basis functions $\phi_i(\cdot)$ are shared while only the linear parameters $\alpha_i$ are specialized to each task. The meta-training phase optimizes the basis function hyperparameters, which are then fixed during meta-testing, and the linear parameters $\alpha_i$ are adapted via Bayesian linear regression, rendering the approach real-time feasible. In the next sections, we will first state the problem formulation and then introduce the overall meta-learning procedure for learning-based MPC. A summary of the method is given in Algorithm~\ref{alg:meta-learning-mpc} .

\subsection{Problem formulation}

We consider dynamic systems of the form
\begin{equation}
  x_{k+1} = g(x_k,u_k) + w_k,
  \label{eq:transition}
\end{equation}
where $x_k \in \mathbb{R}^{n_x}$ is the state variable, $u_k \in \mathbb{R}^{n_u}$ is the input variable, $w_k \sim \mathcal{N}(0, \sigma_w^2 \mathbb{I})$ is i.i.d. additive process noise, and $g$ is the (partially) unknown dynamics of the system.

The system dynamics can be reformulated as
\begin{equation}
  y_k = f(x_k ,u_k) + w_k,
  \label{eq:model}
\end{equation}
where $y_k$ can represent, for instance, state differences $y_k = x_{k+1} - x_k$, or incorporate known dynamics of the system, so that the mapping $f$ represents an unknown residual. This allows for approximating each dimension of $f$ as an independent zero-mean GP~\eqref{eq:gp}, which corresponds to modeling~\eqref{eq:transition} as a collection of GPs with a particular mean function. For the remaining part of the section we will refer to one dimension of the system, but multiple dimensions can be constructed analogously.

%\begin{remark}
% We assume, for simplicity, independence among the dimensions of $f$, and fit to each one a single GP. However, vector-valued GPs allow for approximating the overall vector-valued function by means of matrix kernels. KL expansion for vector-valued GPs... \AC{I need to add Giapi's reference, I think is promising but I need to have a deeper look.}
%\end{remark}

\subsection{Meta-training}

Task data is collected in sets $D_m = \big\{ (x_i,y_i)\big\}_{i=1}^{N_m}$ where $N_m$ is the number of input and output data for each task $m$, and the overall dataset formed by $M$ different tasks is defined as $\mathcal{D}=\{D_m\}_{m=1}^M$. A na{\"i}ve approach would be to train a single GP on each $D_m$, specializing the fit to each task. However, this results in an inefficient use of the available information when the tasks share similarities, described as a distribution $p(m)$ from which each task $m$ is sampled~\cite{CF17}. A more data-efficient approach, which facilitates generalization to a new unseen task, is to train the GP to leverage task similarities and tailor for the re-use of information. We use the approximation in~\eqref{eq:E-dim KL}, and obtain a posterior model description for each task data set $D_m$:
\begin{equation}
  \hat f(x)_{|D_m} = \sum_{i=1}^{E}  \phi_i(x)\alpha_i^m,
  \label{eq:model_qm}
\end{equation}
where $\alpha^m = [ \alpha_1^m, \dots, \alpha_E^m ]^T \sim \mathcal{N}(\mu_\alpha^m, \Sigma_\alpha^m)$, with mean and variance computed via Bayesian linear regression. The main idea is that the basis functions are the same for all $m$, leaving $\alpha^m$ as the only term characterizing a specific task. Finding the best linear approximation~\eqref{eq:model_qm}, such that a finite number of basis functions is shared among tasks, can be cast as an optimization problem over the basis function hyperparameters. For each $m$, we define a variational distribution $q^m(f)$ such that $\hat f(x)_{|D_m} \sim q^m(f)$, i.e.:
\begin{equation}
  q^m(f) \sim \mathcal{N}(\mu_q^m, \Sigma_q^m), \ \mu_q^m = \Phi \mu_\alpha^m, \ \Sigma_q^m = \Phi \Sigma_\alpha^m \Phi^T,
  \label{eq:variational_distribution}
\end{equation}
where $\Phi = [\phi_1(\cdot), \dots, \phi_E(\cdot)]$. We construct a variational bound for the marginal log-likelihood of the GP modeling~\eqref{eq:model}, building on the classical approach for optimizing kernel hyperparameters~\cite{CR17}, and ensuring that the approximate linear model~\eqref{eq:model_qm} is as close as possible to the true underlying GP~\eqref{eq:model}.

We consider the following model likelihood
\begin{equation}
  p(Y,f|X, \theta) = p(Y|f,X,\theta)p(f),
  \label{eq:model_likelihood}
\end{equation}
where $X$ and $Y$ are the overall collection of input and output data, such that $\mathcal{D} = \{X,Y\}$, the likelihood $p(y|f,X,\theta) = \mathcal{N}(f|\sigma_w^2)$ is Gaussian, and the prior $p(f)$ over the latent variable $f$ is a zero-mean GP~\eqref{eq:gp}. Parameter $\theta$ represents the set of basis function hyperparameters that should be optimized. For instance, in the case of a subset of regressors, each basis function is defined by the kernel function evaluated at some input location. Typically, the set of input locations can be chosen a priori as a subset of training data, or test data, but in our formulation, these are considered as hyperparameters in $\theta$, together with the kernel hyperparameters. The model likelihood~\eqref{eq:model_likelihood} is used to construct the following known variational bound~\cite{CB06}:
% \begin{equation}
%   \begin{split}
%     \log p(Y|X,\theta) &= \log \int p(Y,f|X,\theta)df = \log \int q(f) \frac{p(Y,f|X,\theta)}{q(f)}df, \\
%     &\geq  \int q(f) \log\frac{p(Y,f|X,\theta)}{q(f)}df, \text{ (Jensen's inequality)}\\
%     &= \mathbb{E}_q \log\frac{p(Y,f|X,\theta)}{q(f)},
%   \end{split}
%   \label{eq:ELBO}
% \end{equation}
\begin{equation}
  \log p(Y|X,\theta) \geq  \mathbb{E}_q \log\frac{p(Y,f|X,\theta)}{q(f)},
  \label{eq:ELBO}
\end{equation}
where $q(f)$ represents the joint variational distribution $q(f) = \prod_{m=1}^M q^m(f)$. The obtained bound, which is typically referred to as evidence lower bound (ELBO), is maximized in order to approximate the log-marginal likelihood, while obtaining a variational distribution that approximates the true posterior $p(f|Y,X,\theta)$.
% \begin{equation*}
%   \begin{split}
%     0 \leq \log p(Y|X,\theta) - \mathbb{E}_q \log\frac{p(Y,f|X,\theta)}{q(f)} &= \mathbb{E}_q \log \frac{p(Y|X,\theta) q(f)}{p(Y,f|X,\theta)},\\
%     &= \mathbb{E}_q \log \frac{ q(f)}{p(f|Y,X,\theta)},\\
%     &= \text{KL}(q(f) \ || \ p(f|Y,X,\theta)).
%   \end{split}
% \end{equation*}
% \begin{equation*}
%   0 \leq \log p(Y|X,\theta) - \mathbb{E}_q \log\frac{p(Y,f|X,\theta)}{q(f)} = \text{KL}(q(f) \ || \ p(f|Y,X,\theta)).
% \end{equation*}
% This reformulation shows that the bound can be equivalently maximized, by minimizing the KL divergence between the variational distribution $q(f)$ and the true posterior distribution $p(f|Y,X,\theta)$. 
The ELBO~\eqref{eq:ELBO} can be shown to decompose into
\begin{equation}
  \begin{split}
    \mathbb{E}_q \log\frac{p(Y,f|X,\theta)}{q(f)} & = \mathbb{E}_q \log\frac{p(Y|X,f,\theta)p(f)}{q(f)} \\
    & = \mathbb{E}_q \log p(Y|X,f,\theta) + \mathbb{E}_q \log\frac{p(f)}{q(f)} \\
    & = \mathbb{E}_q \log \prod_{m=1}^{M} \prod_{i=1}^{N_m} p(y_i|x_i,f,\theta) + \mathbb{E}_q \log\prod_{m=1}^{M} \frac{p^m(f)}{q^m(f)} \\
    & = \sum_{m=1}^{M}  \Big[ \sum_{i=1}^{N_m} \mathbb{E}_q \log  p(y_i|x_i,f,\theta) +  \mathbb{E}_q \log  \frac{p^m(f)}{q^m(f)} \Big], \\
  \end{split}
  \label{eq:loss_meta_training}
\end{equation}
where $p^m(f)$ is equal to the zero-mean GP prior, with covariance function evaluated at the points $\{x_i\}_{i=1}^{N_m}$. The structure of the obtained loss function, similar to~\cite{SS18}, contains log-likelihood terms \emph{specific} to each task and KL divergence terms, which \emph{share} the same prior distribution, regularizing the variational distribution associated to each task. The expected log-likelihood term can be computed by generating samples from $q^m(f)$, which is a Gaussian distribution itself~\eqref{eq:variational_distribution}, and evaluating the likelihood $p(y_i|x_i,f,\theta) = \mathcal{N}(f,\sigma_w^2)$ at samples of $f$. Furthermore, an analytic form of the KL divergence between two Gaussian distributions is available in terms of mean and variance information. The overall minimization (see Algorithm~\ref{alg:meta-learning-mpc}) can be carried out via gradient descent methods.

\subsection{Meta-testing}

During closed-loop control, the basis functions $\phi_i(\cdot)$ learned during meta-training are fixed while the linear parameters $\alpha_i$ are adapted online as new data is collected during operation via Bayesian linear regression. The adaptive MPC then makes use of the current parameter mean (and potentially also variance) information in the prediction horizon, which is updated at every time-step. We consider a typical MPC formulation in which a control input sequence $U = \{ u_0, \dots, u_{N-1}\}$ is minimized with respect to a cost function $J(U,x_0) = \mathbb{E}_w[\sum_{i=0}^{N} \ l_k(x_k,u_k)] $, where $N$ is the length of the control horizon, and $l_k: \mathbb{R}^{n_x} \times \mathbb{R}^{n_u} \rightarrow \mathbb{R}$ is a potentially time-varying stage cost function. The optimization problem is subject to the model dynamics and state or input constraints. At each time-step, only the first input of the optimal sequence is applied to the system, i.e. the MPC controller is implemented in a receding-horizon fashion.

\begin{algorithm}[H]
  \caption{Meta Learning MPC}
  \label{alg:meta-learning-mpc}
  \begin{algorithmic}[1]
    \State Collect task data $\mathcal{D}$ using a nominal controller for tasks $m \sim p(m)$ for meta-training. Fix number of basis functions $E$.
    \State \textbf{Meta-training} (offline):
    \State Set the basis functions $\phi_i(\cdot)$, and initialize hyperparameters $\theta$.
    \State Optimize $\theta$ for all tasks in $\mathcal{D}$:
    $$ \min_{\theta} -\sum_{m=1}^{M}  \Big[ \sum_{i=1}^{N_m} \mathbb{E}_q \log  p(y_i|x_i,f,\theta) +  \mathbb{E}_q \log  \frac{p^m(f)}{q^m(f)} \Big]
    $$

    \State \textbf{Meta-testing} (online):
    \State Fix hyperparameters $\theta$ from meta-training.
    \State Initialize prior mean and variance of $\alpha^0 = [\alpha^0_1, \dots, \alpha^0_E] $.
    \For { $t \gets 1$ to $T$}
    \State Solve MPC problem using current residual model $f^t = \Phi\alpha^t$.
    \State Apply input and obtain new measurement $y^t$.
    \State Update $\alpha^t$ via Bayesian linear regression.
    \EndFor

  \end{algorithmic}
\end{algorithm}

\section{Results}

In the following, we first present an illustrative example of the well-known mountain-car problem, which is used for describing the mechanisms of the proposed meta-learning approach. As a second example, we consider the control of an autonomous miniature race car, where the goal is to enable adaptation to unseen road conditions via the proposed meta-learning MPC approach. The details of the examples and the implementations are provided in the supplementary.

\subsection{Mountain-car problem}

\begin{figure}[H]
  \hspace{-1cm}
  \begin{subfigure}[t]{0.35\textwidth}
    \input{train_traj.tex}
    \caption{Meta-training trajectories}
    \label{fg:train_traj}
  \end{subfigure}%
  \begin{subfigure}[t]{0.35\textwidth}
    \input{test_traj.tex}
    \caption{Meta-testing trajectories}
    \label{fg:test_traj}
  \end{subfigure}%
  \begin{subfigure}[t]{0.35\textwidth}
    \input{elbo.tex}
    \caption{Negative ELBO using cosine basis function}
    \label{fg:elbo}
  \end{subfigure}
  \caption{Meta-training results for known basis function of the residual.}
\end{figure}
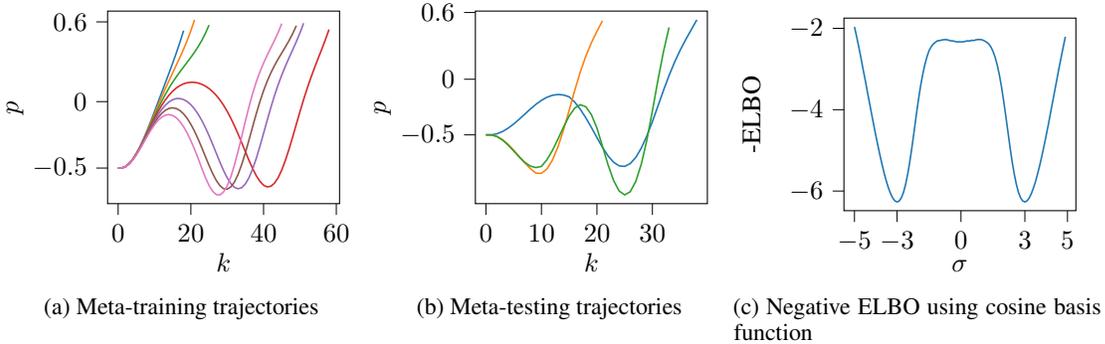

The mountain-car problem is typically used as a benchmark for RL, and considers the problem of driving from a valley past the top of a hill. The dynamics are given by~\cite{RS18}:
\begin{equation*}
  x_{k+1} = \begin{bmatrix}
    p_{k+1} \\ v_{k+1}
  \end{bmatrix} = \begin{bmatrix}
    p_{k} + T_sv_{k} \\ v_{k} - T_s \cos(3p_k)\theta_1 + T_su_k\theta_2
  \end{bmatrix} + \begin{bmatrix}
    0 \\ w_k
  \end{bmatrix},
\end{equation*}
where $p$ is the position of the car, $v$ is the velocity, $u$ corresponds to the acceleration and the additive disturbance is $w_k \sim \mathcal{N}(0,0.001^2)$. The sampling time is $T_s = 0.2 s$ and the actuator gain is set as $\theta_2 = 0.3$. The residual dynamics to be learned is given by  $y_k = v_{k+1} - v_{k} - T_su_k\theta_2 = -T_s \cos(3p_k)\theta_1 + w_k$, and is therefore linear in the parameter $\theta_1$. The basis function can be directly defined as $\tilde \phi(p) = -T_s \cos(3p)$, and the parameter determining each task is $\theta_1$. Varying parameter $\theta_1$ corresponds to changing the slope of the mountain, and in turn defines a task. The training trajectories, depicted in Figure~\ref{fg:train_traj}, are generated by a nominal MPC controller with perfect knowledge of the overall dynamics, with $\theta_1 \in D_{\theta_1}^{train} = \{0.3,0.35,0.4,0.45,0.5,0.55,0.6\}$. The test tasks, depicted in Figure~\ref{fg:test_traj}, are similarly generated, with $\theta_1 \in D_{\theta_1}^{test}=\{0.65,0.9,1.3\}$, and are used as ground truth for comparisons.

We first consider the case of a known basis function $\tilde \phi(p)$, and with hyperparameter represented by the cosine frequency $\sigma$. Figure~\ref{fg:elbo} shows the associated (negative) ELBO~\eqref{eq:ELBO}, which displays two clear minima at $\sigma = \pm 3$, which is expected given symmetry of the cosine.

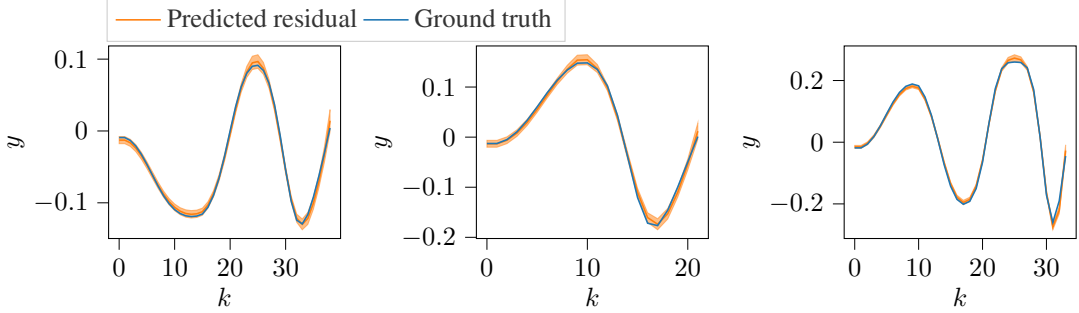
\begin{figure}[H]
  \hspace{-1cm}
  \begin{subfigure}[t]{0.35\textwidth}
    %\centering
    \input{MC7.tex}
  \end{subfigure}%
  \begin{subfigure}[t]{0.35\textwidth}
    %\centering
    \input{MC8.tex}
  \end{subfigure}%
  \begin{subfigure}[t]{0.35\textwidth}
    %\centering
    \input{MC9.tex}
  \end{subfigure}
  \vspace{-0.5cm}
  \caption{Comparison between ground truth and predicted residual dynamics, for which the mean is highlighted in bold. The shaded area around the mean is bounded by $\pm2\sigma$ confidence intervals: (left) $\theta_1 = 0.65$, (center) $\theta_1 = 0.9$, (right) $\theta_1=1.3$.}
  \label{fg:meta-testing}
\end{figure}

\vspace{-0.6cm}
We then consider $\tilde \phi(p)$ unknown, and carry out the proposed meta-learning MPC scheme. The basis functions chosen for this example are a subset of regressors, which consist in evaluating the kernel function at a finite number of input locations $E$, so that $\phi_i(\cdot) = K(\tilde p^i, \cdot), \ i=1,\dots,E$, where $E=4$. The considered kernel is the squared-exponential, so that the overall basis function hyperparameters are $\theta = \{\lambda, \sigma^2_w, \sigma^2 , \{\tilde p^i \}_{i=1}^E\}$, corresponding to lengthscale, output noise, scaling factor and input locations respectively. In this example, we focus on investigating learning quality rather than the control performance, showing in Figure~\eqref{fg:meta-testing} the residual dynamics learned using previously collected meta-testing data, which is considered as ground truth. Closed-loop adaptation is further investigated in the proposed miniature race car example.

\subsection{Miniature race car}
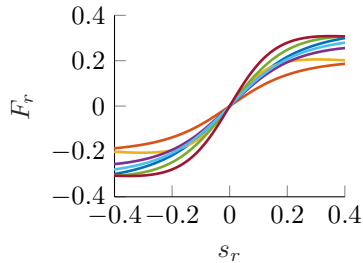
\begin{figure}[H]
  \hspace{-0.5cm}
  \centering
  \input{pacejka.tex}
  \caption{Pacejka models for the $7$ different meta-training tasks.}
  \label{fg:pacejka}
\end{figure}

In this example, we consider the problem of controlling a miniature race car modeled by a kinematic bicycle model~\cite{RR11}, to race on a track with different road conditions.
%The model encompasses $6$ state variables and $2$ inputs. The dynamics equations are
% \begin{equation}
%   \label{eq:car_dynamics}
%   \begin{split}
%     &x_{k+1} = x_k v_{x_k} \cos(\psi_k) - v_{y_k} \sin(\psi_k)\\
%     &y_{k+1} = v_{x_k} \sin(\psi_k) + v_{y_k} \cos(\psi_k)\\
%     &\psi_{k+1} = \omega_k\\
%     &v_{x_k} = \frac{1}{m} \left( F_{x_k} - F_{f_k} \sin(\delta_k) + mv_{y_k} \omega_k \right) \\
%     &v_{y_k} = \frac{1}{m} \left(F_{r_k} + F_{f_k} \cos(\delta_k) - mv_{x_k} \omega_k \right)\\
%     &\omega_{k+1} = \frac{1}{I_z}\left( F_{f_k} l_f \cos(\delta_k) - F_{r_k} l_r \right),\\
%   \end{split}
% \end{equation}
% where $x$ and $y$ are the coordinates of the car, $\psi$ and $\omega$ are the yaw angle and yaw rate, and $v_x$ and $v_y$ the longitudinal and lateral velocity, respectively. The steering angle is denoted by $\delta$ and is an input variable for the system.  The car parameters $m$, $I_z$, $l_f$, and $l_r$ represent the mass, the inertia on the z-axis with respect to the car frame, the distance between the center of mass of the car and the front wheel axis, and the distance between the center of mass of the car and the rear wheel axis. The longitudinal force is denoted by $F_x$ and is modeled by the following equation
% \begin{equation}
%   \label{eq:drivetrain}
%   F_x = C_m T - C_r - C_d v_x^2,
% \end{equation}
% where the first term is the drive-train force and $T$, the torque applied by the motor, is the other input of the system. The second and third terms in~\eqref{eq:drivetrain} are the rolling friction and aerodynamic drag, respectively. 
Front and rear tire lateral forces, $F_f$ and $F_r$, are modeled using a simplified Pacejka model~\cite{HP15}
\begin{equation}
  \begin{split}
    F_{r} &= D_r \sin(C_r \arctan(B_r s_r)) \\
    F_{f} &= D_f \sin(C_f \arctan(B_f s_f)),
  \end{split}
  \label{eq:forces}
\end{equation}
where $s_r$ and $s_f$ are the slip angles, which depend on the current velocity of the car.
% \begin{align}\label{eq:slip}
%   \begin{aligned}
%     s_r & = \arctan\left( \frac{v_y - \omega l_r}{v_x} \right)           \\
%     s_f & = \arctan\left( \frac{v_y + \omega l_f}{v_x} \right) - \delta.
%   \end{aligned}
% \end{align}
The Pacejka coefficients $B_f, C_f, D_f, B_r, C_r,$ and $D_r$ are usually identified from data, can vary over time, and depend on specific road conditions, e.g., driving on a dry or wet surface. In this example, we use meta-learning to learn online the Pacejka front and rear tire models leveraging previous tasks, by approximating~\eqref{eq:forces} as $F_{i} = \Phi(s_{i})\alpha, \ i=r,f$. We collect meta-training data for $7$ tasks with different tire models, illustrated in Figure~\ref{fg:pacejka}, where every data set consists of $200$ points.
%The paramater $B$ is spanning between $0.20$ and $0.37$, $C$ between $1.06$ and $0.16$, and $D$ between $3.32$ and $5.31$. 
\vspace{-0.4cm}
We feed the training data to the proposed meta-training algorithm using again a subset of regressors, with $E=14$. After the meta-training phase, meta-testing is performed in closed-loop using an adaptive formulation of the model predictive contouring controller (MPCC)~\cite{DL10}, where the vector of parameters $\alpha = [\alpha_1, \dots, \alpha_E]$ is adapted online via Bayesian linear regression at each time step, making use of the mean sequential update \footnote{In the case of Gaussian distributions, the posterior mean update obtained via Bayesian linear regression corresponds to the maximum a posteriori (MAP) estimate~\cite{CB06}. Furthermore, it can be shown that the MAP estimate corresponds to a maximum likelihood estimate with Ridge regularization, which justifies the regularizing term in~\eqref{eq:sequential_update}.}, i.e.,
\begin{equation}
  \mu_{\alpha_{k+1}} = \mu_{\alpha_{k}} + \eta \left( (y_k - \mu_{\alpha_{k}}^T \Phi(x_k)) \Phi(x_k) + \sigma_w^2 \mu_{\alpha_{k}}  \right),
  \label{eq:sequential_update}
\end{equation}
where $\eta= 0.0005$ is the learning rate, and $\sigma_w=0.02$ is the noise standard deviation.

% \begin{table}[h]
%   \centering
%   \begin{tabular}{ccccccccc}
%     m     & $I_z$     & $l_r$ & $l_f$ & $C_m$  & $C_r$  & $C_d$  & $\eta$ & $\sigma_w$ \\ \hline
%     0.041 & 0.0000278 & 0.033 & 0.029 & 0.2443 & 0.0358 & 0.0019 & 0.0001 & 0.02
%   \end{tabular}
%   \caption{\label{table:parameters}Parameters used in the simulation.}
% \end{table}

\begin{figure}[H]
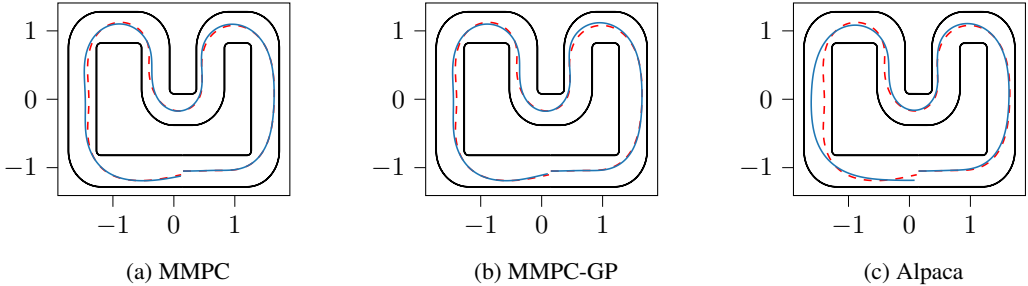

  \  \begin{subfigure}[t]{0.35\textwidth}
    %\centering
    \input{car_lap_ml.tex}
    \caption{MMPC}
    \label{fig:car_meta_testing_mmpc}
  \end{subfigure}%
  \begin{subfigure}[t]{0.35\textwidth}
    %\centering
    \input{car_lap_fullgp.tex}
    \caption{MMPC-GP}
    \label{fig:car_meta_testing_mmpc_gp}
  \end{subfigure}%
  \begin{subfigure}[t]{0.35\textwidth}
    %\centering
    \input{car_lap_alpaca.tex}
    \caption{Alpaca}
    \label{fig:car_meta_testing_alpaca}
  \end{subfigure}
  \vspace{0cm}
  \caption{Counter-clockwise trajectory using a ground truth model predictive contouring controller (dashed red line) compared to MMPC (\ref{fig:car_meta_testing_mmpc}), MMPC-GP (\ref{fig:car_meta_testing_mmpc_gp}), and Alpaca (\ref{fig:car_meta_testing_alpaca}).}
  \label{fig:car_meta_testing}
\end{figure}

% The simulation was performed in a ROS environment using both C++ and Python code. In particular, we simulated three nodes: the car model running at $120$Hz, the state estimator running at $250$Hz, and the controller running at $30$Hz. The meta-training is implemented in Python using GPyTorch~\cite{JG18} and Adams. The solver of the model predictive contouring controller was generated using in Python ACADOS~\cite{RV19} and called from a C++ ROS node for performance reasons. The simulation was performed on an Intel Core i7 3.3GHz machine with $32$ GB of RAM.
The meta-testing results are shown in Figure~\ref{fg:rmse} and in Figure~\ref{fig:car_meta_testing}. We compare our proposed meta-learning MPC (MMPC) against a variation, which we refer to as MMPC-GP, which specifies the variational distribution $q^m(f)$ as an exact GP, and extracts the basis functions a-posteriori, thus optimizing only the kernel hyperparameters during meta-training. The approaches are compared against Alpaca~\cite{JH18a},~\cite{JH18b}, where the basis functions are designed as neural networks, adapting the last linear layer online. The network weights are meta-trained together with the prior mean and variance over the last layer, by optimizing the posterior predictive distribution. The performance of all methods is comparably good already after the first lap, shown for one noise realization in Figure~\ref{fig:car_meta_testing}, which is quantified in terms of the cumulative root mean squared error (RMSE) with respect to a ground truth MPCC in Figure~\ref{fg:rmse} . The neural network for Alpaca was chosen trading-off the model complexity against computational requirements for enabling closed-loop control. While the RSME for Alpaca is only slightly higher, it has significantly more parameters to be tuned: in this example, considering weights and biases of the neural network, these are around $\sim100$ times more than for the proposed approach. MMPC achieves lower RMSE while reducing the number of parameters, which becomes essential for embedded platforms with limited computation power and memory storage. The difference in performance w.r.t. MMPC-GP is minimal and likely due to the lack of input location optimization during meta-training. Furthermore, while this does not pose a real constraint given that meta-training is performed offline, the hyperparameter optimization using basis functions, rather than exact GPs, is consistently faster while providing the same performance in this example. Finally, we tested MMPC with changing grip conditions, where the grip decreases by about $36\%$ in the (orange) second half of the track. In Figure~\ref{fg:adaptation}, the resulting trajectory shows fast adaptation to the new grip, and a performance that is very close to the result of the ground truth MPCC controller which has perfect knowledge of the changing grip.

\vspace{-0.5cm}
\begin{figure}[h]
  \hspace{-0.5cm}
  \begin{subfigure}[t]{0.5\textwidth}
    \centering
    \input{car_lap_rmse.tex}
    \caption{Comparison of cumulative RMSE for 1000 noise realizations along the first lap.}
    \label{fg:rmse}
  \end{subfigure}
  \hfill
  \begin{subfigure}[t]{0.5\textwidth}
    \centering
    \input{car_lap_adaptaion.tex}
    \caption{Counter-clockwise trajectory using a ground truth MPCC ($\mbox{- - -}$) compared to MMPC (---).}
    \label{fg:adaptation}
  \end{subfigure}
  \caption[caption for subfigures]{}
\end{figure}

\section{Conclusion}

We presented a meta-learning approach for data-efficient adaptive MPC, making use of linear GP approximations via basis functions. We provide both the ability to generalize by extracting shared information from related task data, and an efficient inference scheme by applying Bayesian linear regression to tailor the linear combination of basis functions online to the task. Results are shown in simulation examples for the mountain-car problem and for a miniature race car adapting to changing road conditions. The presented performance, both in terms of low cumulative errors with respect to a ground truth trajectory, and in terms of computational tractability, motivates future experimental work in this direction.

% \section*{Broader Impact}
% There has been recent interest in quantifying the environmental impact of using machine learning algorithms~\cite{Wired}, \cite{Guardian}, particularly concerning the energy consumed for training with large amounts of data. Meta-learning provides an important step toward an efficient use of energy and data, by "recycling" information. Furthermore, we address the important challenge of using learning-based control techniques for the many applications that require embedded systems, and therefore need computationally and data-efficient methods.

% An important limitation, shared by most of the proposed meta-learning approaches in the literature, is given by the underlying assumption that the training and test tasks should be related by a shared underlying probability distribution, meaning that data cannot be randomly collected. There exist, however, many application domains in which relations among tasks are quite natural, for example in the area of manufacturing or machine tools. 

% \begin{ack}
%   This work was supported by ...
% \end{ack}

\bibliographystyle{abbrvnat}
\bibliography{neurips_2020}

\end{document}

%% file: train_traj.tex
% This file was created by tikzplotlib v0.9.2.
\begin{tikzpicture}

\definecolor{color0}{rgb}{0.12156862745098,0.466666666666667,0.705882352941177}
\definecolor{color1}{rgb}{1,0.498039215686275,0.0549019607843137}
\definecolor{color2}{rgb}{0.172549019607843,0.627450980392157,0.172549019607843}
\definecolor{color3}{rgb}{0.83921568627451,0.152941176470588,0.156862745098039}
\definecolor{color4}{rgb}{0.580392156862745,0.403921568627451,0.741176470588235}
\definecolor{color5}{rgb}{0.549019607843137,0.337254901960784,0.294117647058824}
\definecolor{color6}{rgb}{0.890196078431372,0.466666666666667,0.76078431372549}

\begin{axis}[
tick align=outside,
tick pos=left,
x grid style={white!69.0196078431373!black},
xlabel={\(\displaystyle k\)},
xmin=-2.9, xmax=60.9,
xtick style={color=black},
y grid style={white!69.0196078431373!black},
ylabel={\(\displaystyle p\)},
ymin=-0.767720348236776, ymax=0.681161393627198,
ytick style={color=black},
ytick = {-0.5, 0 , 0.6},
scale = {0.45}
]
\addplot [semithick, color0]
table {%
0 -0.5
1 -0.5
2 -0.488848805427551
3 -0.466546535491943
4 -0.433493137359619
5 -0.390483617782593
6 -0.338678598403931
7 -0.279539465904236
8 -0.214720726013184
9 -0.145923733711243
10 -0.0747216939926147
12 0.0702461004257202
14 0.21578049659729
15 0.289766311645508
16 0.366179943084717
17 0.446849346160889
18 0.534059762954712
};
\addplot [semithick, color1]
table {%
0 -0.5
1 -0.5
2 -0.488990306854248
3 -0.466971039772034
4 -0.434402942657471
5 -0.392202138900757
6 -0.341703057289124
7 -0.284580826759338
8 -0.222724795341492
9 -0.158069014549255
11 -0.0271912813186646
12 0.036554217338562
13 0.098346471786499
14 0.158222794532776
17 0.333617448806763
18 0.394994020462036
19 0.460816264152527
20 0.53337025642395
21 0.615303158760071
};
\addplot [semithick, color2]
table {%
0 -0.5
1 -0.5
2 -0.489131450653076
3 -0.467394471168518
4 -0.435309171676636
5 -0.393908500671387
6 -0.344696402549744
7 -0.289552807807922
8 -0.230590224266052
10 -0.109654307365417
11 -0.0513119697570801
12 0.00388801097869873
13 0.0552771091461182
14 0.102666974067688
15 0.146276712417603
16 0.186639189720154
17 0.224518179893494
18 0.260840177536011
19 0.296656250953674
20 0.333126068115234
21 0.371525049209595
22 0.413270711898804
23 0.459967374801636
24 0.513464570045471
25 0.575926065444946
};
\addplot [semithick, color3]
table {%
0 -0.5
1 -0.5
2 -0.489272832870483
3 -0.467819452285767
4 -0.436216592788696
5 -0.395611882209778
6 -0.347671747207642
7 -0.294474601745605
8 -0.238342523574829
9 -0.181631684303284
10 -0.126512050628662
11 -0.0747860670089722
12 -0.0277791023254395
13 0.013679027557373
14 0.0491998195648193
15 0.0787355899810791
16 0.102466821670532
17 0.120697975158691
18 0.133772730827332
19 0.142014741897583
20 0.145686745643616
21 0.144967794418335
22 0.139940738677979
23 0.130589127540588
24 0.116800904273987
25 0.0983765125274658
26 0.075046181678772
27 0.0464936494827271
28 0.0123952627182007
29 -0.0275282859802246
30 -0.073439359664917
31 -0.125289082527161
32 -0.182703375816345
33 -0.244860887527466
34 -0.310381412506104
35 -0.377259731292725
36 -0.442882537841797
37 -0.504156351089478
38 -0.557746291160583
39 -0.600385665893555
40 -0.629184246063232
41 -0.641872644424438
42 -0.636953830718994
43 -0.613781213760376
44 -0.572604656219482
45 -0.514617443084717
46 -0.441993117332458
47 -0.357854127883911
48 -0.266077756881714
49 -0.170887470245361
50 -0.0762605667114258
51 0.0146803855895996
52 0.100090146064758
53 0.179517507553101
54 0.253750324249268
55 0.324530839920044
56 0.394279718399048
57 0.465906739234924
58 0.542724132537842
};
\addplot [semithick, color4]
table {%
0 -0.5
1 -0.5
2 -0.48941445350647
3 -0.468243718147278
4 -0.437120795249939
5 -0.397303819656372
6 -0.350617647171021
7 -0.299329280853271
9 -0.193054795265198
10 -0.142947435379028
11 -0.0975785255432129
12 -0.0583982467651367
13 -0.0263668298721313
14 -0.0020294189453125
15 0.0143705606460571
16 0.0227710008621216
17 0.023189902305603
18 0.0156551599502563
19 0.000169038772583008
20 -0.0232951641082764
21 -0.0547595024108887
22 -0.094174861907959
23 -0.141320943832397
24 -0.195674180984497
25 -0.256256937980652
26 -0.321491599082947
27 -0.38910174369812
28 -0.456108808517456
29 -0.518968343734741
30 -0.573849678039551
31 -0.617009043693542
32 -0.645164608955383
33 -0.655788660049438
34 -0.647279262542725
35 -0.619044542312622
36 -0.571558237075806
37 -0.506422758102417
38 -0.426419734954834
39 -0.335446834564209
40 -0.238222718238831
41 -0.139697432518005
42 -0.0442781448364258
43 0.0448722839355469
44 0.126199007034302
45 0.199706792831421
46 0.266631126403809
47 0.329038619995117
48 0.389510154724121
49 0.45095956325531
50 0.516578793525696
51 0.589874029159546
};
\addplot [semithick, color5]
table {%
0 -0.5
1 -0.5
2 -0.489556193351746
3 -0.468668699264526
4 -0.438024401664734
5 -0.398989200592041
6 -0.353540182113647
7 -0.30412483215332
8 -0.253452777862549
9 -0.20424211025238
10 -0.158972263336182
11 -0.11970055103302
12 -0.087973952293396
13 -0.0648441314697266
14 -0.0509523153305054
15 -0.0466457605361938
16 -0.0520830154418945
17 -0.0673047304153442
18 -0.0922590494155884
19 -0.126766443252563
20 -0.170436143875122
21 -0.222534418106079
22 -0.281818866729736
23 -0.346379995346069
24 -0.413535952568054
25 -0.479844689369202
26 -0.541286587715149
27 -0.593608140945435
28 -0.632762908935547
29 -0.655330300331116
30 -0.658820748329163
31 -0.641841650009155
32 -0.604180693626404
33 -0.546878337860107
34 -0.472308874130249
35 -0.384204149246216
36 -0.287471413612366
37 -0.187672853469849
38 -0.0901875495910645
39 0.000693202018737793
40 0.0823739767074585
41 0.154054880142212
42 0.21640419960022
43 0.271062135696411
44 0.320195198059082
45 0.366210222244263
46 0.411618232727051
47 0.459020376205444
48 0.511170148849487
49 0.57108473777771
};
\addplot [semithick, color6]
table {%
0 -0.5
1 -0.5
2 -0.489697694778442
3 -0.469093203544617
4 -0.43892502784729
5 -0.400663614273071
6 -0.356433033943176
8 -0.26081371307373
9 -0.215186953544617
10 -0.174580812454224
11 -0.141144871711731
12 -0.116491794586182
13 -0.101718783378601
14 -0.0974950790405273
15 -0.104162335395813
16 -0.1218101978302
17 -0.150295734405518
18 -0.189196109771729
19 -0.237697958946228
20 -0.294436573982239
21 -0.357327222824097
22 -0.427476644515991
23 -0.501101016998291
24 -0.571146965026855
25 -0.63081169128418
26 -0.675064563751221
27 -0.699730277061462
28 -0.701862096786499
29 -0.679894208908081
30 -0.633694410324097
31 -0.564649343490601
32 -0.475820899009705
33 -0.372044324874878
34 -0.259696006774902
35 -0.14588737487793
36 -0.0371559858322144
37 0.0618373155593872
38 0.148979425430298
39 0.224533319473267
40 0.290444850921631
41 0.349598407745361
43 0.461036682128906
44 0.520432591438293
45 0.587350606918335
};
\end{axis}

\end{tikzpicture}

%% file: test_traj.tex
% This file was created by tikzplotlib v0.9.2.
\begin{tikzpicture}

\definecolor{color0}{rgb}{0.12156862745098,0.466666666666667,0.705882352941177}
\definecolor{color1}{rgb}{1,0.498039215686275,0.0549019607843137}
\definecolor{color2}{rgb}{0.172549019607843,0.627450980392157,0.172549019607843}

\begin{axis}[
tick align=outside,
tick pos=left,
x grid style={white!69.0196078431373!black},
xlabel={\(\displaystyle k\)},
xmin=-1.9, xmax=39.9,
xtick style={color=black},
y grid style={white!69.0196078431373!black},
ylabel={\(\displaystyle p\)},
ymin=-1.1183713847531, ymax=0.612798738828023,
ytick style={color=black},
ytick = {-0.5, 0 , 0.6},
scale = {0.45}
]
\addplot [semithick, color0]
table {%
0 -0.5
1 -0.5
2 -0.489839196205139
3 -0.469517469406128
4 -0.439824461936951
5 -0.402331233024597
6 -0.359303832054138
8 -0.268054723739624
9 -0.225903272628784
10 -0.189787983894348
11 -0.161926984786987
12 -0.143964052200317
13 -0.136993050575256
14 -0.141634702682495
15 -0.158111572265625
16 -0.205410957336426
17 -0.269808530807495
18 -0.346798658370972
19 -0.433187127113342
20 -0.524106979370117
21 -0.612279176712036
22 -0.688411712646484
23 -0.745708465576172
24 -0.778667211532593
25 -0.783555030822754
26 -0.758432865142822
27 -0.703027844429016
28 -0.618783712387085
29 -0.509210348129272
30 -0.380313038825989
31 -0.240537405014038
32 -0.0995967388153076
33 0.0338246822357178
34 0.154398322105408
35 0.261105537414551
36 0.356552362442017
38 0.534109115600586
};
\addplot [semithick, color1]
table {%
0 -0.5
1 -0.5
2 -0.514546394348145
3 -0.543639659881592
4 -0.585710525512695
5 -0.637618064880371
7 -0.752018690109253
8 -0.803488850593567
9 -0.844175577163696
10 -0.846062898635864
11 -0.806423425674438
12 -0.725141286849976
13 -0.604849100112915
14 -0.452091932296753
15 -0.278646469116211
16 -0.100864768028259
17 0.0647797584533691
18 0.208059787750244
19 0.328017711639404
20 0.430763602256775
21 0.525578022003174
};
\addplot [semithick, color2]
table {%
0 -0.5
1 -0.5
2 -0.515678286552429
3 -0.547034978866577
4 -0.591626882553101
5 -0.64456582069397
6 -0.698965668678284
7 -0.746906042098999
8 -0.780734062194824
9 -0.794274091720581
10 -0.783562898635864
11 -0.723115921020508
12 -0.638102412223816
13 -0.535789728164673
14 -0.427963733673096
15 -0.330236434936523
16 -0.259223937988281
17 -0.228712201118469
18 -0.24725341796875
19 -0.318027257919312
20 -0.439139485359192
21 -0.602326631546021
22 -0.768107771873474
23 -0.909721374511719
24 -1.00452148914337
25 -1.03968179225922
26 -1.01126778125763
27 -0.918866872787476
28 -0.762767553329468
29 -0.546474695205688
30 -0.283991098403931
31 -0.00594139099121094
32 0.249866724014282
33 0.465683102607727
};
\end{axis}

\end{tikzpicture}

%% file: elbo.tex
% This file was created by tikzplotlib v0.9.2.
\begin{tikzpicture}

\definecolor{color0}{rgb}{0.12156862745098,0.466666666666667,0.705882352941177}

\begin{axis}[
tick align=outside,
tick pos=left,
x grid style={white!69.0196078431373!black},
xlabel={\(\displaystyle \sigma\)},
xmin=-5.495, xmax=5.39499999999996,
xtick style={color=black},
extra x ticks = {-3,3},
y grid style={white!69.0196078431373!black},
ylabel={-ELBO},
ymin=-6.48242296710477, ymax=-1.74867120699798,
ytick style={color=black},
scale = {0.45}
]
\addplot [semithick, color0]
table {%
-5 -1.96384179592133
-4.90000009536743 -2.19930863380432
-4.80000019073486 -2.45181274414062
-4.69999980926514 -2.71278882026672
-4.59999990463257 -2.99499225616455
-4.5 -3.25826740264893
-4.40000009536743 -3.54332780838013
-4.09999990463257 -4.36692094802856
-4 -4.63115692138672
-3.90000009536743 -4.88292694091797
-3.79999995231628 -5.12670564651489
-3.70000004768372 -5.35866546630859
-3.59999990463257 -5.57257080078125
-3.5 -5.76388597488403
-3.40000009536743 -5.93101453781128
-3.29999995231628 -6.06785154342651
-3.20000004768372 -6.17376899719238
-3.09999990463257 -6.24262094497681
-3 -6.26725244522095
-2.90000009536743 -6.24516725540161
-2.79999995231628 -6.16616487503052
-2.70000004768372 -6.03396844863892
-2.59999990463257 -5.8366003036499
-2.5 -5.57962322235107
-2.40000009536743 -5.26861095428467
-2.29999995231628 -4.91561031341553
-2.20000004768372 -4.53330945968628
-2.09999990463257 -4.13859891891479
-2 -3.76603484153748
-1.89999997615814 -3.4292414188385
-1.79999995231628 -3.14235591888428
-1.70000004768372 -2.89822030067444
-1.60000002384186 -2.70540237426758
-1.5 -2.54792451858521
-1.39999997615814 -2.45519948005676
-1.29999995231628 -2.38360595703125
-1.10000002384186 -2.30386114120483
-1 -2.30206704139709
-0.899999976158142 -2.2873067855835
-0.799999952316284 -2.27834248542786
-0.700000047683716 -2.27733206748962
-0.600000023841858 -2.28722858428955
-0.5 -2.30519366264343
-0.399999976158142 -2.30777525901794
-0.299999952316284 -2.31863570213318
-0.200000047683716 -2.33515095710754
-0.100000023841858 -2.32970213890076
-0 -2.33331513404846
0.100000023841858 -2.33305406570435
0.299999952316284 -2.30705142021179
0.399999976158142 -2.31282138824463
0.5 -2.29494118690491
0.600000023841858 -2.29992365837097
0.700000047683716 -2.28349590301514
0.799999952316284 -2.27406859397888
0.899999976158142 -2.27508139610291
1 -2.28400182723999
1.10000002384186 -2.3076229095459
1.20000004768372 -2.3405864238739
1.29999995231628 -2.38196992874146
1.39999997615814 -2.45371031761169
1.5 -2.56077790260315
1.60000002384186 -2.70592284202576
1.70000004768372 -2.89673709869385
1.79999995231628 -3.13804912567139
1.89999997615814 -3.4319953918457
2 -3.77085542678833
2.09999990463257 -4.13967943191528
2.29999995231628 -4.91160106658936
2.40000009536743 -5.26999521255493
2.5 -5.57808065414429
2.59999990463257 -5.83362627029419
2.70000004768372 -6.0320873260498
2.79999995231628 -6.16777276992798
2.90000009536743 -6.24408102035522
3 -6.26710987091064
3.09999990463257 -6.24200677871704
3.20000004768372 -6.17400789260864
3.29999995231628 -6.06707954406738
3.40000009536743 -5.92835474014282
3.5 -5.764328956604
3.59999990463257 -5.57232856750488
3.70000004768372 -5.36279201507568
3.79999995231628 -5.13236093521118
3.90000009536743 -4.8876805305481
4 -4.63402414321899
4.09999990463257 -4.37223148345947
4.19999980926514 -4.08982753753662
4.5 -3.26452827453613
4.59999990463257 -2.98217296600342
4.80000019073486 -2.46097683906555
4.90000009536743 -2.20602369308472
};
\end{axis}

\end{tikzpicture}

%% file: MC7.tex
% This file was created by tikzplotlib v0.9.2.
\begin{tikzpicture}

\definecolor{color0}{rgb}{1,0.498039215686275,0.0549019607843137}
\definecolor{color1}{rgb}{0.12156862745098,0.466666666666667,0.705882352941177}

\begin{axis}[
legend cell align={left},
legend columns = {-1},
legend style={fill opacity=0.8, draw opacity=1, text opacity=1, draw=white!80!black,at={(1.97,1.25)}},
tick align=outside,
tick pos=left,
x grid style={white!69.0196078431373!black},
xlabel={\(\displaystyle k\)},
xmin=-1.9, xmax=39.9,
xtick style={color=black},
y grid style={white!69.0196078431373!black},
ylabel={\(\displaystyle y\)},
ymin=-0.149696872829866, ymax=0.118342630211956,
ytick style={color=black},
scale = {0.45}
]
\path [draw=color0, fill=color0, opacity=0.5]
(axis cs:0,-0.0176934113987723)
--(axis cs:0,-0.00774693550851833)
--(axis cs:1,-0.00774693550851833)
--(axis cs:2,-0.0115085852100466)
--(axis cs:3,-0.0189531371543579)
--(axis cs:4,-0.0296176807451813)
--(axis cs:5,-0.0426530028789921)
--(axis cs:6,-0.0568812297027849)
--(axis cs:7,-0.0709518555634907)
--(axis cs:8,-0.0836024712305118)
--(axis cs:9,-0.0939489056302851)
--(axis cs:10,-0.101644372343625)
--(axis cs:11,-0.106800700059818)
--(axis cs:12,-0.109754205144426)
--(axis cs:13,-0.110820956587328)
--(axis cs:14,-0.110115630638709)
--(axis cs:15,-0.107452582276263)
--(axis cs:16,-0.0984529568766365)
--(axis cs:17,-0.0831419588765054)
--(axis cs:18,-0.0608440679526839)
--(axis cs:19,-0.0319628085620866)
--(axis cs:20,0.00127287133206969)
--(axis cs:21,0.035264805308972)
--(axis cs:22,0.0659651852225836)
--(axis cs:23,0.089976566407402)
--(axis cs:24,0.104059025883199)
--(axis cs:25,0.106159016437327)
--(axis cs:26,0.0953952050728324)
--(axis cs:27,0.0720214349955085)
--(axis cs:28,0.0378344582141406)
--(axis cs:29,-0.00431606586956529)
--(axis cs:30,-0.0500423294260084)
--(axis cs:31,-0.0905177384804182)
--(axis cs:32,-0.115774999105898)
--(axis cs:33,-0.122549054203971)
--(axis cs:34,-0.113282327724942)
--(axis cs:35,-0.0920381389286758)
--(axis cs:36,-0.0615041017094143)
--(axis cs:37,-0.0216234239981291)
--(axis cs:38,0.0297519450947845)
--(axis cs:38,-0.00127086103780749)
--(axis cs:38,-0.00127086103780749)
--(axis cs:37,-0.0430465370200479)
--(axis cs:36,-0.0789049893415726)
--(axis cs:35,-0.108959022031545)
--(axis cs:34,-0.130099995180929)
--(axis cs:33,-0.137513259055238)
--(axis cs:32,-0.127129916931395)
--(axis cs:31,-0.0992874834045191)
--(axis cs:30,-0.0593681854964198)
--(axis cs:29,-0.0142686328320511)
--(axis cs:28,0.0276426296709138)
--(axis cs:27,0.0593236119023476)
--(axis cs:26,0.0787871714194195)
--(axis cs:25,0.0871453645179063)
--(axis cs:24,0.0855434988503454)
--(axis cs:23,0.074438241419354)
--(axis cs:22,0.053977089488559)
--(axis cs:21,0.0251368345519473)
--(axis cs:20,-0.0086806608013865)
--(axis cs:19,-0.0416813128888851)
--(axis cs:18,-0.0698896618475312)
--(axis cs:17,-0.0918244571009575)
--(axis cs:16,-0.107549290855694)
--(axis cs:15,-0.117362228520674)
--(axis cs:14,-0.120393563687751)
--(axis cs:13,-0.121209059453446)
--(axis cs:12,-0.119977855504805)
--(axis cs:11,-0.116630611086781)
--(axis cs:10,-0.110964692229308)
--(axis cs:9,-0.102824522420568)
--(axis cs:8,-0.0922857921823661)
--(axis cs:7,-0.0797634841103783)
--(axis cs:6,-0.0660296373191114)
--(axis cs:5,-0.0521578072368701)
--(axis cs:4,-0.0393741798301042)
--(axis cs:3,-0.0288384141060523)
--(axis cs:2,-0.0214421627653018)
--(axis cs:1,-0.0176934113987723)
--(axis cs:0,-0.0176934113987723)
--cycle;

\addplot [semithick, color0]
table {%
0 -0.0127202272415161
1 -0.0127202272415161
2 -0.0164753198623657
3 -0.0238957405090332
4 -0.0344959497451782
5 -0.0474053621292114
6 -0.0614554882049561
7 -0.0753576755523682
8 -0.0879441499710083
9 -0.0983867645263672
10 -0.106304526329041
11 -0.11171567440033
12 -0.114866018295288
13 -0.116014957427979
14 -0.115254640579224
15 -0.112407445907593
16 -0.103001117706299
17 -0.0874831676483154
18 -0.0653668642044067
19 -0.0368220806121826
20 -0.00370395183563232
21 0.0302008390426636
22 0.0599710941314697
23 0.0822074413299561
24 0.0948013067245483
25 0.0966521501541138
26 0.0870912075042725
27 0.0656725168228149
28 0.0327385663986206
29 -0.0092923641204834
30 -0.0547052621841431
31 -0.0949026346206665
32 -0.121452450752258
33 -0.130031108856201
34 -0.121691107749939
35 -0.100498557090759
36 -0.070204496383667
37 -0.0323349237442017
38 0.0142405033111572
};
\addlegendentry{Predicted residual}
\addplot [semithick, color1]
table {%
0 -0.00919604301452637
1 -0.00919520854949951
2 -0.0131435394287109
3 -0.0209989547729492
4 -0.0323294401168823
5 -0.0462583303451538
6 -0.0615110397338867
7 -0.0766098499298096
8 -0.0901813507080078
9 -0.101271152496338
10 -0.109490633010864
11 -0.114959359169006
12 -0.11806309223175
13 -0.119176268577576
14 -0.11844003200531
15 -0.115647435188293
16 -0.106089234352112
17 -0.0896890163421631
18 -0.0657647848129272
19 -0.0348297357559204
20 0.000198125839233398
21 0.0341790914535522
22 0.0616905689239502
23 0.0803544521331787
24 0.0900493860244751
25 0.0914144515991211
26 0.0841953754425049
27 0.0666462182998657
28 0.0366201400756836
29 -0.00560832023620605
30 -0.0541751384735107
31 -0.0975968837738037
32 -0.124238014221191
33 -0.129332304000854
34 -0.116302013397217
35 -0.092115044593811
36 -0.0624545812606812
37 -0.0301520824432373
38 0.00409770011901855
};
\addlegendentry{Ground truth}
\end{axis}

\end{tikzpicture}

%% file: MC8.tex
% This file was created by tikzplotlib v0.9.2.
\begin{tikzpicture}

\definecolor{color0}{rgb}{1,0.498039215686275,0.0549019607843137}
\definecolor{color1}{rgb}{0.12156862745098,0.466666666666667,0.705882352941177}

\begin{axis}[
legend cell align={left},
legend style={fill opacity=0.8, draw opacity=1, text opacity=1, draw=white!80!black},
tick align=outside,
tick pos=left,
x grid style={white!69.0196078431373!black},
xlabel={\(\displaystyle k\)},
xmin=-1.05, xmax=22.05,
xtick style={color=black},
y grid style={white!69.0196078431373!black},
ylabel={\(\displaystyle y\)},
ymin=-0.202377239930142, ymax=0.182010771883195,
ytick style={color=black},
scale = {0.45}
]
\path [draw=color0, fill=color0, opacity=0.5]
(axis cs:0,-0.0201247315099693)
--(axis cs:0,-0.00637689999001518)
--(axis cs:1,-0.00637689999001518)
--(axis cs:2,0.000646980903954501)
--(axis cs:3,0.0148060403645539)
--(axis cs:4,0.0354474445062796)
--(axis cs:5,0.061020918661952)
--(axis cs:6,0.089276927296515)
--(axis cs:7,0.11767542238994)
--(axis cs:8,0.14338581130554)
--(axis cs:9,0.163606445607746)
--(axis cs:10,0.164538589528043)
--(axis cs:11,0.144850169714062)
--(axis cs:12,0.104292112481936)
--(axis cs:13,0.0448702429847293)
--(axis cs:14,-0.0291288205821242)
--(axis cs:15,-0.10225706953483)
--(axis cs:16,-0.150987715314286)
--(axis cs:17,-0.16288514753254)
--(axis cs:18,-0.142464363873756)
--(axis cs:19,-0.100655982581347)
--(axis cs:20,-0.0437099240988702)
--(axis cs:21,0.0281327301529169)
--(axis cs:21,-0.00447794229871341)
--(axis cs:21,-0.00447794229871341)
--(axis cs:20,-0.0655440210650465)
--(axis cs:19,-0.119717500007325)
--(axis cs:18,-0.163580319693207)
--(axis cs:17,-0.18490505757499)
--(axis cs:16,-0.170360101651587)
--(axis cs:15,-0.118162856532861)
--(axis cs:14,-0.0433859356980567)
--(axis cs:13,0.0327574678973071)
--(axis cs:12,0.0926111999694695)
--(axis cs:11,0.12871211953341)
--(axis cs:10,0.144452682613308)
--(axis cs:9,0.143732473087411)
--(axis cs:8,0.127496224596782)
--(axis cs:7,0.105081957474248)
--(axis cs:6,0.0780305157714489)
--(axis cs:5,0.0494302406722898)
--(axis cs:4,0.0230027942685987)
--(axis cs:3,0.00166313765277635)
--(axis cs:2,-0.0129150363642479)
--(axis cs:1,-0.0201247315099693)
--(axis cs:0,-0.0201247315099693)
--cycle;

\addplot [semithick, color0]
table {%
0 -0.0132508277893066
1 -0.0132508277893066
2 -0.006134033203125
3 0.00823462009429932
4 0.0292251110076904
5 0.0552256107330322
6 0.0836536884307861
7 0.11137866973877
8 0.135441064834595
9 0.153669476509094
10 0.154495596885681
11 0.136781096458435
12 0.0984516143798828
13 0.0388138294219971
14 -0.0362573862075806
15 -0.110209941864014
16 -0.160673856735229
17 -0.173895120620728
18 -0.153022289276123
19 -0.110186696052551
20 -0.0546269416809082
21 0.0118273496627808
};
\addplot [semithick, color1]
table {%
0 -0.0127322673797607
1 -0.0127332210540771
2 -0.00488924980163574
3 0.0108176469802856
4 0.033347487449646
5 0.0603759288787842
6 0.0884627103805542
7 0.113917708396912
8 0.133996725082397
9 0.14763343334198
10 0.148214221000671
11 0.135050177574158
12 0.10232400894165
13 0.0434424877166748
14 -0.0383192300796509
15 -0.120686173439026
16 -0.171822071075439
17 -0.176610827445984
18 -0.146059989929199
19 -0.099657416343689
20 -0.0494858026504517
21 0.001068115234375
};
\end{axis}

\end{tikzpicture}

%% file: MC9.tex
% This file was created by tikzplotlib v0.9.2.
\begin{tikzpicture}

\definecolor{color0}{rgb}{1,0.498039215686275,0.0549019607843137}
\definecolor{color1}{rgb}{0.12156862745098,0.466666666666667,0.705882352941177}

\begin{axis}[
legend cell align={left},
legend style={fill opacity=0.8, draw opacity=1, text opacity=1, at={(0.03,0.97)}, anchor=north west, draw=white!80!black},
tick align=outside,
tick pos=left,
x grid style={white!69.0196078431373!black},
xlabel={\(\displaystyle k\)},
xmin=-1.65, xmax=34.65,
xtick style={color=black},
y grid style={white!69.0196078431373!black},
ylabel={\(\displaystyle y\)},
ymin=-0.312225861485973, ymax=0.312086679645922,
ytick style={color=black},
scale = {0.45}
]
\path [draw=color0, fill=color0, opacity=0.5]
(axis cs:0,-0.0195813551564588)
--(axis cs:0,-0.0103619038302886)
--(axis cs:1,-0.0103619038302886)
--(axis cs:2,0.00107892892993608)
--(axis cs:3,0.0239300735460747)
--(axis cs:4,0.0560827681105092)
--(axis cs:5,0.0931742398949082)
--(axis cs:6,0.129396796428661)
--(axis cs:7,0.159234624484196)
--(axis cs:8,0.178918728856946)
--(axis cs:9,0.18645503165855)
--(axis cs:10,0.180509782079565)
--(axis cs:11,0.144697215961631)
--(axis cs:12,0.0887300379622269)
--(axis cs:13,0.0157469883416583)
--(axis cs:14,-0.0623820398868575)
--(axis cs:15,-0.128835388520815)
--(axis cs:16,-0.171826405009368)
--(axis cs:17,-0.18846811487898)
--(axis cs:18,-0.178499886814387)
--(axis cs:19,-0.136604077371596)
--(axis cs:20,-0.0544172016746608)
--(axis cs:21,0.0636940476282438)
--(axis cs:22,0.171712146627748)
--(axis cs:23,0.242068878701665)
--(axis cs:24,0.274873836285001)
--(axis cs:25,0.283708836867199)
--(axis cs:26,0.276715431173)
--(axis cs:27,0.245771098607558)
--(axis cs:28,0.1686132954298)
--(axis cs:29,0.0235228972053249)
--(axis cs:30,-0.157462882868409)
--(axis cs:31,-0.263701559688374)
--(axis cs:32,-0.209694534239327)
--(axis cs:33,-0.00863089358521205)
--(axis cs:33,-0.0456113084493706)
--(axis cs:33,-0.0456113084493706)
--(axis cs:32,-0.232263403224055)
--(axis cs:31,-0.283848018707251)
--(axis cs:30,-0.169805935009065)
--(axis cs:29,0.0141996916278295)
--(axis cs:28,0.158878143014268)
--(axis cs:27,0.232922945954415)
--(axis cs:26,0.257435648291954)
--(axis cs:25,0.261707288664344)
--(axis cs:24,0.256191044707837)
--(axis cs:23,0.22964695971461)
--(axis cs:22,0.161966521258064)
--(axis cs:21,0.054163225686129)
--(axis cs:20,-0.0638128379968999)
--(axis cs:19,-0.14801814848022)
--(axis cs:18,-0.191950449300455)
--(axis cs:17,-0.202502658723103)
--(axis cs:16,-0.184907259738089)
--(axis cs:15,-0.13994716491071)
--(axis cs:14,-0.0718659392316651)
--(axis cs:13,0.00645819936709138)
--(axis cs:12,0.0790924976140733)
--(axis cs:11,0.134986478267891)
--(axis cs:10,0.17071663790114)
--(axis cs:9,0.176609901536759)
--(axis cs:8,0.169136481421536)
--(axis cs:7,0.149517689722238)
--(axis cs:6,0.119687291215852)
--(axis cs:5,0.0835223175657197)
--(axis cs:4,0.0465907176340271)
--(axis cs:3,0.0146049517993659)
--(axis cs:2,-0.00816011181261172)
--(axis cs:1,-0.0195813551564588)
--(axis cs:0,-0.0195813551564588)
--cycle;

\addplot [semithick, color0]
table {%
0 -0.0149716138839722
1 -0.0149716138839722
2 -0.00354063510894775
3 0.0192675590515137
4 0.0513367652893066
5 0.0883482694625854
6 0.124541997909546
7 0.154376149177551
8 0.174027562141418
9 0.181532502174377
10 0.175613164901733
11 0.139841794967651
12 0.0839112997055054
13 0.011102557182312
14 -0.0671240091323853
15 -0.134391307830811
16 -0.178366780281067
17 -0.195485353469849
18 -0.185225129127502
19 -0.142311096191406
20 -0.0591150522232056
21 0.0589286088943481
22 0.166839361190796
23 0.235857963562012
24 0.265532493591309
25 0.272708058357239
26 0.267075538635254
27 0.239346981048584
28 0.163745760917664
29 0.0188612937927246
30 -0.163634419441223
31 -0.273774862289429
32 -0.220978975296021
33 -0.0271210670471191
};
\addplot [semithick, color1]
table {%
0 -0.0183912515640259
1 -0.0183923244476318
2 -0.00617611408233643
3 0.0182657241821289
4 0.0526944398880005
5 0.0922980308532715
6 0.130561232566833
7 0.161439895629883
8 0.181256771087646
9 0.188678026199341
10 0.182833433151245
11 0.146495819091797
12 0.0875654220581055
13 0.00950801372528076
14 -0.0735750198364258
15 -0.14250385761261
16 -0.185263991355896
17 -0.201163530349731
18 -0.191692113876343
19 -0.150374412536621
20 -0.065176248550415
21 0.0608377456665039
22 0.174067258834839
23 0.238198757171631
24 0.257872104644775
25 0.259934663772583
26 0.258491158485413
27 0.240968227386475
28 0.17095422744751
29 0.0178300142288208
30 -0.171208024024963
31 -0.259958267211914
32 -0.190309643745422
33 -0.0449473857879639
};
\end{axis}

\end{tikzpicture}

%% file: pacejka.tex
\definecolor{mycolor1}{rgb}{0.00000,0.44700,0.74100}%
\definecolor{mycolor2}{rgb}{0.85000,0.32500,0.09800}%
\definecolor{mycolor3}{rgb}{0.92900,0.69400,0.12500}%
\definecolor{mycolor4}{rgb}{0.49400,0.18400,0.55600}%
\definecolor{mycolor5}{rgb}{0.46600,0.67400,0.18800}%
\definecolor{mycolor6}{rgb}{0.30100,0.74500,0.93300}%
\definecolor{mycolor7}{rgb}{0.63500,0.07800,0.18400}%
\begin{tikzpicture}

	\begin{axis}[%
			width=6.028in,
			height=4.754in,
			at={(1.011in,0.642in)},
			scale only axis,
			xmin=-0.4,
			xmax=0.4,
			xlabel style={font=\color{white!15!black}},
			xlabel={\(\displaystyle s_r\)},
			xtick={-0.4, -0.2, 0,0.2,0.4},
			ymin=-0.4,
			ymax=0.4,
			ylabel style={font=\color{white!15!black}},
			ylabel={\(\displaystyle F_r\)},
			axis background/.style={fill=white},
			axis x line*=bottom,
			axis y line*=left,
			%xmajorgrids,
			%ymajorgrids,
			legend style={legend cell align=left, align=left, draw=white!15!black},
			scale={0.2}
		]
		\addplot [color=mycolor1, line width=1pt]
		table[row sep=crcr]{%
				-1	-0.343271828414858\\
				-0.99	-0.343093449839139\\
				-0.98	-0.342909062792256\\
				-0.97	-0.342718422881722\\
				-0.96	-0.342521273813198\\
				-0.95	-0.342317346714816\\
				-0.94	-0.342106359417908\\
				-0.93	-0.341888015690982\\
				-0.92	-0.341662004423597\\
				-0.91	-0.341427998756491\\
				-0.9	-0.341185655154034\\
				-0.89	-0.340934612414779\\
				-0.88	-0.340674490615536\\
				-0.87	-0.340404889984044\\
				-0.86	-0.340125389694895\\
				-0.85	-0.339835546582982\\
				-0.84	-0.339534893768218\\
				-0.83	-0.33922293918484\\
				-0.82	-0.338899164008008\\
				-0.81	-0.338563020969863\\
				-0.8	-0.338213932556546\\
				-0.79	-0.337851289077008\\
				-0.78	-0.337474446593682\\
				-0.77	-0.337082724704283\\
				-0.76	-0.336675404163145\\
				-0.75	-0.33625172432955\\
				-0.74	-0.335810880429484\\
				-0.73	-0.335352020616187\\
				-0.72	-0.334874242813661\\
				-0.71	-0.334376591326054\\
				-0.7	-0.333858053194477\\
				-0.69	-0.33331755428136\\
				-0.68	-0.332753955060913\\
				-0.67	-0.332166046092601\\
				-0.66	-0.331552543152802\\
				-0.65	-0.330912081997965\\
				-0.64	-0.33024321273066\\
				-0.63	-0.329544393737856\\
				-0.62	-0.328813985168686\\
				-0.61	-0.328050241916783\\
				-0.6	-0.327251306070072\\
				-0.59	-0.326415198788686\\
				-0.58	-0.325539811569521\\
				-0.57	-0.324622896853854\\
				-0.56	-0.323662057932522\\
				-0.55	-0.322654738101491\\
				-0.54	-0.32159820901933\\
				-0.53	-0.320489558217286\\
				-0.52	-0.31932567571253\\
				-0.51	-0.318103239675879\\
				-0.5	-0.316818701107249\\
				-0.49	-0.315468267475465\\
				-0.48	-0.314047885284391\\
				-0.47	-0.312553221534973\\
				-0.46	-0.310979644063463\\
				-0.45	-0.309322200750372\\
				-0.44	-0.307575597613492\\
				-0.43	-0.305734175822656\\
				-0.42	-0.303791887704859\\
				-0.41	-0.301742271847373\\
				-0.4	-0.299578427455148\\
				-0.39	-0.297292988178835\\
				-0.38	-0.294878095703347\\
				-0.37	-0.292325373476209\\
				-0.36	-0.289625901062503\\
				-0.35	-0.286770189741699\\
				-0.34	-0.283748160113682\\
				-0.33	-0.280549122659489\\
				-0.32	-0.277161762408939\\
				-0.31	-0.273574129104096\\
				-0.3	-0.26977363451499\\
				-0.29	-0.265747058861203\\
				-0.28	-0.261480568616314\\
				-0.27	-0.256959748315122\\
				-0.26	-0.252169649334748\\
				-0.25	-0.247094858963156\\
				-0.24	-0.241719593378081\\
				-0.23	-0.236027818402709\\
				-0.22	-0.230003402038741\\
				-0.21	-0.223630302748473\\
				-0.2	-0.216892797200247\\
				-0.19	-0.209775750631303\\
				-0.18	-0.202264932037585\\
				-0.17	-0.194347374990458\\
				-0.16	-0.18601178293385\\
				-0.15	-0.177248975283656\\
				-0.14	-0.168052367525965\\
				-0.13	-0.158418474844298\\
				-0.12	-0.14834742473214\\
				-0.11	-0.137843459798838\\
				-0.1	-0.126915407896878\\
				-0.09	-0.11557709323412\\
				-0.08	-0.10384765981397\\
				-0.07	-0.0917517779287001\\
				-0.0599999999999999	-0.0793197060319005\\
				-0.0499999999999999	-0.0665871845129062\\
				-0.04	-0.053595144828284\\
				-0.03	-0.0403892269263096\\
				-0.02	-0.0270191093678622\\
				-0.01	-0.0135376690734918\\
				0	0\\
				0.01	0.0135376690734918\\
				0.02	0.0270191093678622\\
				0.03	0.0403892269263096\\
				0.04	0.053595144828284\\
				0.0499999999999999	0.0665871845129062\\
				0.0599999999999999	0.0793197060319005\\
				0.07	0.0917517779287001\\
				0.08	0.10384765981397\\
				0.09	0.11557709323412\\
				0.1	0.126915407896878\\
				0.11	0.137843459798838\\
				0.12	0.14834742473214\\
				0.13	0.158418474844298\\
				0.14	0.168052367525965\\
				0.15	0.177248975283656\\
				0.16	0.18601178293385\\
				0.17	0.194347374990458\\
				0.18	0.202264932037585\\
				0.19	0.209775750631303\\
				0.2	0.216892797200247\\
				0.21	0.223630302748473\\
				0.22	0.230003402038741\\
				0.23	0.236027818402709\\
				0.24	0.241719593378081\\
				0.25	0.247094858963156\\
				0.26	0.252169649334748\\
				0.27	0.256959748315122\\
				0.28	0.261480568616314\\
				0.29	0.265747058861203\\
				0.3	0.26977363451499\\
				0.31	0.273574129104096\\
				0.32	0.277161762408939\\
				0.33	0.280549122659489\\
				0.34	0.283748160113682\\
				0.35	0.286770189741699\\
				0.36	0.289625901062503\\
				0.37	0.292325373476209\\
				0.38	0.294878095703347\\
				0.39	0.297292988178835\\
				0.4	0.299578427455148\\
				0.41	0.301742271847373\\
				0.42	0.303791887704859\\
				0.43	0.305734175822656\\
				0.44	0.307575597613492\\
				0.45	0.309322200750372\\
				0.46	0.310979644063463\\
				0.47	0.312553221534973\\
				0.48	0.314047885284391\\
				0.49	0.315468267475465\\
				0.5	0.316818701107249\\
				0.51	0.318103239675879\\
				0.52	0.31932567571253\\
				0.53	0.320489558217286\\
				0.54	0.32159820901933\\
				0.55	0.322654738101491\\
				0.56	0.323662057932522\\
				0.57	0.324622896853854\\
				0.58	0.325539811569521\\
				0.59	0.326415198788686\\
				0.6	0.327251306070072\\
				0.61	0.328050241916783\\
				0.62	0.328813985168686\\
				0.63	0.329544393737856\\
				0.64	0.33024321273066\\
				0.65	0.330912081997965\\
				0.66	0.331552543152802\\
				0.67	0.332166046092601\\
				0.68	0.332753955060913\\
				0.69	0.33331755428136\\
				0.7	0.333858053194477\\
				0.71	0.334376591326054\\
				0.72	0.334874242813661\\
				0.73	0.335352020616187\\
				0.74	0.335810880429484\\
				0.75	0.33625172432955\\
				0.76	0.336675404163145\\
				0.77	0.337082724704283\\
				0.78	0.337474446593682\\
				0.79	0.337851289077008\\
				0.8	0.338213932556546\\
				0.81	0.338563020969863\\
				0.82	0.338899164008008\\
				0.83	0.33922293918484\\
				0.84	0.339534893768218\\
				0.85	0.339835546582982\\
				0.86	0.340125389694895\\
				0.87	0.340404889984044\\
				0.88	0.340674490615536\\
				0.89	0.340934612414779\\
				0.9	0.341185655154034\\
				0.91	0.341427998756491\\
				0.92	0.341662004423597\\
				0.93	0.341888015690982\\
				0.94	0.342106359417908\\
				0.95	0.342317346714816\\
				0.96	0.342521273813198\\
				0.97	0.342718422881722\\
				0.98	0.342909062792256\\
				0.99	0.343093449839139\\
				1	0.343271828414858\\
			};

		\addplot [color=mycolor2, line width=1pt]
		table[row sep=crcr]{%
				-1	-0.205259223335118\\
				-0.99	-0.205228439031084\\
				-0.98	-0.205195168552966\\
				-0.97	-0.205159286292916\\
				-0.96	-0.205120659843809\\
				-0.95	-0.205079149586944\\
				-0.94	-0.205034608251851\\
				-0.93	-0.204986880446118\\
				-0.92	-0.204935802152972\\
				-0.91	-0.204881200194217\\
				-0.9	-0.20482289165587\\
				-0.89	-0.204760683273669\\
				-0.88	-0.204694370775379\\
				-0.87	-0.204623738176548\\
				-0.86	-0.204548557026132\\
				-0.85	-0.204468585598062\\
				-0.84	-0.204383568024537\\
				-0.83	-0.204293233366452\\
				-0.82	-0.204197294615993\\
				-0.81	-0.204095447626015\\
				-0.8	-0.203987369960362\\
				-0.79	-0.203872719658801\\
				-0.78	-0.203751133909695\\
				-0.77	-0.203622227622982\\
				-0.76	-0.203485591895372\\
				-0.75	-0.203340792359016\\
				-0.74	-0.203187367404143\\
				-0.73	-0.203024826265381\\
				-0.72	-0.202852646960589\\
				-0.71	-0.202670274070134\\
				-0.7	-0.202477116343494\\
				-0.69	-0.202272544119031\\
				-0.68	-0.202055886541585\\
				-0.67	-0.201826428561314\\
				-0.66	-0.201583407695872\\
				-0.65	-0.201326010536604\\
				-0.64	-0.201053368977918\\
				-0.63	-0.200764556147459\\
				-0.62	-0.200458582012977\\
				-0.61	-0.200134388640124\\
				-0.6	-0.199790845073572\\
				-0.59	-0.199426741812032\\
				-0.58	-0.199040784845925\\
				-0.57	-0.198631589224599\\
				-0.56	-0.198197672118254\\
				-0.55	-0.197737445338055\\
				-0.54	-0.197249207276492\\
				-0.53	-0.196731134228834\\
				-0.52	-0.196181271055802\\
				-0.51	-0.195597521147291\\
				-0.5	-0.194977635647554\\
				-0.49	-0.194319201903638\\
				-0.48	-0.193619631101618\\
				-0.47	-0.192876145059385\\
				-0.46	-0.192085762151007\\
				-0.45	-0.191245282346399\\
				-0.44	-0.190351271361826\\
				-0.43	-0.189400043932373\\
				-0.42	-0.18838764623772\\
				-0.41	-0.187309837538471\\
				-0.4	-0.18616207111293\\
				-0.39	-0.184939474625038\\
				-0.38	-0.183636830104715\\
				-0.37	-0.182248553783695\\
				-0.36	-0.180768676105146\\
				-0.35	-0.179190822315884\\
				-0.34	-0.177508194157876\\
				-0.33	-0.175713553303282\\
				-0.32	-0.173799207326154\\
				-0.31	-0.171756999175904\\
				-0.3	-0.169578301313399\\
				-0.29	-0.167254015889857\\
				-0.28	-0.164774582589583\\
				-0.27	-0.16212999601573\\
				-0.26	-0.159309834766053\\
				-0.25	-0.156303304611351\\
				-0.24	-0.153099298435736\\
				-0.23	-0.149686475801471\\
				-0.22	-0.146053365130482\\
				-0.21	-0.142188491510269\\
				-0.2	-0.138080532985476\\
				-0.19	-0.133718507831954\\
				-0.18	-0.129091994666329\\
				-0.17	-0.124191386258581\\
				-0.16	-0.119008176531509\\
				-0.15	-0.113535278408964\\
				-0.14	-0.107767367903007\\
				-0.13	-0.101701247141569\\
				-0.12	-0.0953362160262169\\
				-0.11	-0.0886744390419237\\
				-0.1	-0.0817212906680225\\
				-0.09	-0.0744856601931567\\
				-0.08	-0.0669801949114754\\
				-0.07	-0.0592214600940819\\
				-0.0599999999999999	-0.051229995181363\\
				-0.0499999999999999	-0.0430302486209659\\
				-0.04	-0.034650378798865\\
				-0.03	-0.0261219154495668\\
				-0.02	-0.0174792843756162\\
				-0.01	-0.00875920756864169\\
				0	0\\
				0.01	0.00875920756864169\\
				0.02	0.0174792843756162\\
				0.03	0.0261219154495668\\
				0.04	0.034650378798865\\
				0.0499999999999999	0.0430302486209659\\
				0.0599999999999999	0.051229995181363\\
				0.07	0.0592214600940819\\
				0.08	0.0669801949114754\\
				0.09	0.0744856601931567\\
				0.1	0.0817212906680225\\
				0.11	0.0886744390419237\\
				0.12	0.0953362160262169\\
				0.13	0.101701247141569\\
				0.14	0.107767367903007\\
				0.15	0.113535278408964\\
				0.16	0.119008176531509\\
				0.17	0.124191386258581\\
				0.18	0.129091994666329\\
				0.19	0.133718507831954\\
				0.2	0.138080532985476\\
				0.21	0.142188491510269\\
				0.22	0.146053365130482\\
				0.23	0.149686475801471\\
				0.24	0.153099298435736\\
				0.25	0.156303304611351\\
				0.26	0.159309834766053\\
				0.27	0.16212999601573\\
				0.28	0.164774582589583\\
				0.29	0.167254015889857\\
				0.3	0.169578301313399\\
				0.31	0.171756999175904\\
				0.32	0.173799207326154\\
				0.33	0.175713553303282\\
				0.34	0.177508194157876\\
				0.35	0.179190822315884\\
				0.36	0.180768676105146\\
				0.37	0.182248553783695\\
				0.38	0.183636830104715\\
				0.39	0.184939474625038\\
				0.4	0.18616207111293\\
				0.41	0.187309837538471\\
				0.42	0.18838764623772\\
				0.43	0.189400043932373\\
				0.44	0.190351271361826\\
				0.45	0.191245282346399\\
				0.46	0.192085762151007\\
				0.47	0.192876145059385\\
				0.48	0.193619631101618\\
				0.49	0.194319201903638\\
				0.5	0.194977635647554\\
				0.51	0.195597521147291\\
				0.52	0.196181271055802\\
				0.53	0.196731134228834\\
				0.54	0.197249207276492\\
				0.55	0.197737445338055\\
				0.56	0.198197672118254\\
				0.57	0.198631589224599\\
				0.58	0.199040784845925\\
				0.59	0.199426741812032\\
				0.6	0.199790845073572\\
				0.61	0.200134388640124\\
				0.62	0.200458582012977\\
				0.63	0.200764556147459\\
				0.64	0.201053368977918\\
				0.65	0.201326010536604\\
				0.66	0.201583407695872\\
				0.67	0.201826428561314\\
				0.68	0.202055886541585\\
				0.69	0.202272544119031\\
				0.7	0.202477116343494\\
				0.71	0.202670274070134\\
				0.72	0.202852646960589\\
				0.73	0.203024826265381\\
				0.74	0.203187367404143\\
				0.75	0.203340792359016\\
				0.76	0.203485591895372\\
				0.77	0.203622227622982\\
				0.78	0.203751133909695\\
				0.79	0.203872719658801\\
				0.8	0.203987369960362\\
				0.81	0.204095447626015\\
				0.82	0.204197294615993\\
				0.83	0.204293233366452\\
				0.84	0.204383568024537\\
				0.85	0.204468585598062\\
				0.86	0.204548557026132\\
				0.87	0.204623738176548\\
				0.88	0.204694370775379\\
				0.89	0.204760683273669\\
				0.9	0.20482289165587\\
				0.91	0.204881200194217\\
				0.92	0.204935802152972\\
				0.93	0.204986880446118\\
				0.94	0.205034608251851\\
				0.95	0.205079149586944\\
				0.96	0.205120659843809\\
				0.97	0.205159286292916\\
				0.98	0.205195168552966\\
				0.99	0.205228439031084\\
				1	0.205259223335118\\
			};

		\addplot [color=mycolor3, line width=1pt]
		table[row sep=crcr]{%
				-1	-0.167820816116773\\
				-0.99	-0.168188080276723\\
				-0.98	-0.168560872791146\\
				-0.97	-0.168939300902342\\
				-0.96	-0.1693234736425\\
				-0.95	-0.169713501790973\\
				-0.94	-0.170109497820687\\
				-0.93	-0.170511575832285\\
				-0.92	-0.170919851474515\\
				-0.91	-0.171334441849135\\
				-0.9	-0.171755465398444\\
				-0.89	-0.172183041773321\\
				-0.88	-0.172617291679387\\
				-0.87	-0.17305833669865\\
				-0.86	-0.17350629908367\\
				-0.85	-0.173961301520913\\
				-0.84	-0.174423466859616\\
				-0.83	-0.174892917801976\\
				-0.82	-0.17536977655005\\
				-0.81	-0.175854164404138\\
				-0.8	-0.176346201306823\\
				-0.79	-0.176846005326126\\
				-0.78	-0.17735369207044\\
				-0.77	-0.177869374026995\\
				-0.76	-0.178393159814622\\
				-0.75	-0.178925153340411\\
				-0.74	-0.179465452848574\\
				-0.73	-0.180014149848386\\
				-0.72	-0.180571327906394\\
				-0.71	-0.181137061286249\\
				-0.7	-0.181711413417372\\
				-0.69	-0.182294435171307\\
				-0.68	-0.182886162921858\\
				-0.67	-0.183486616362078\\
				-0.66	-0.184095796047635\\
				-0.65	-0.184713680632167\\
				-0.64	-0.185340223755682\\
				-0.63	-0.185975350541998\\
				-0.62	-0.186618953655346\\
				-0.61	-0.187270888859675\\
				-0.6	-0.187930970016672\\
				-0.59	-0.188598963449918\\
				-0.58	-0.189274581592897\\
				-0.57	-0.189957475827431\\
				-0.56	-0.190647228406545\\
				-0.55	-0.191343343341375\\
				-0.54	-0.192045236115408\\
				-0.53	-0.192752222070806\\
				-0.52	-0.193463503290464\\
				-0.51	-0.194178153775576\\
				-0.5	-0.194895102691381\\
				-0.49	-0.195613115423181\\
				-0.48	-0.196330772150136\\
				-0.47	-0.19704644360554\\
				-0.46	-0.197758263648692\\
				-0.45	-0.198464098224894\\
				-0.44	-0.199161510236164\\
				-0.43	-0.199847719785797\\
				-0.42	-0.200519559195012\\
				-0.41	-0.201173422119767\\
				-0.4	-0.201805206021335\\
				-0.39	-0.202410247166504\\
				-0.38	-0.202983247254755\\
				-0.37	-0.20351819069359\\
				-0.36	-0.20400825147456\\
				-0.35	-0.204445688548536\\
				-0.34	-0.204821728569568\\
				-0.33	-0.20512643488627\\
				-0.32	-0.20534856172718\\
				-0.31	-0.205475392677497\\
				-0.3	-0.205492562812915\\
				-0.29	-0.205383864286052\\
				-0.28	-0.20513103580884\\
				-0.27	-0.20471353741156\\
				-0.26	-0.204108313173821\\
				-0.25	-0.203289546420163\\
				-0.24	-0.202228414274952\\
				-0.23	-0.20089285161159\\
				-0.22	-0.199247338445392\\
				-0.21	-0.197252729825847\\
				-0.2	-0.194866153351379\\
				-0.19	-0.192041006528887\\
				-0.18	-0.188727094132308\\
				-0.17	-0.184870954015073\\
				-0.16	-0.180416427652163\\
				-0.15	-0.175305537665061\\
				-0.14	-0.169479736723323\\
				-0.13	-0.162881587844382\\
				-0.12	-0.155456921964505\\
				-0.11	-0.147157491215492\\
				-0.1	-0.137944092549696\\
				-0.09	-0.127790074723865\\
				-0.08	-0.11668506378507\\
				-0.07	-0.104638654436221\\
				-0.0599999999999999	-0.0916837292735729\\
				-0.0499999999999999	-0.0778790030618458\\
				-0.04	-0.0633103668315346\\
				-0.03	-0.0480906474141901\\
				-0.02	-0.0323575147293245\\
				-0.01	-0.0162694589574552\\
				0	0\\
				0.01	0.0162694589574552\\
				0.02	0.0323575147293245\\
				0.03	0.0480906474141901\\
				0.04	0.0633103668315346\\
				0.0499999999999999	0.0778790030618458\\
				0.0599999999999999	0.0916837292735729\\
				0.07	0.104638654436221\\
				0.08	0.11668506378507\\
				0.09	0.127790074723865\\
				0.1	0.137944092549696\\
				0.11	0.147157491215492\\
				0.12	0.155456921964505\\
				0.13	0.162881587844382\\
				0.14	0.169479736723323\\
				0.15	0.175305537665061\\
				0.16	0.180416427652163\\
				0.17	0.184870954015073\\
				0.18	0.188727094132308\\
				0.19	0.192041006528887\\
				0.2	0.194866153351379\\
				0.21	0.197252729825847\\
				0.22	0.199247338445392\\
				0.23	0.20089285161159\\
				0.24	0.202228414274952\\
				0.25	0.203289546420163\\
				0.26	0.204108313173821\\
				0.27	0.20471353741156\\
				0.28	0.20513103580884\\
				0.29	0.205383864286052\\
				0.3	0.205492562812915\\
				0.31	0.205475392677497\\
				0.32	0.20534856172718\\
				0.33	0.20512643488627\\
				0.34	0.204821728569568\\
				0.35	0.204445688548536\\
				0.36	0.20400825147456\\
				0.37	0.20351819069359\\
				0.38	0.202983247254755\\
				0.39	0.202410247166504\\
				0.4	0.201805206021335\\
				0.41	0.201173422119767\\
				0.42	0.200519559195012\\
				0.43	0.199847719785797\\
				0.44	0.199161510236164\\
				0.45	0.198464098224894\\
				0.46	0.197758263648692\\
				0.47	0.19704644360554\\
				0.48	0.196330772150136\\
				0.49	0.195613115423181\\
				0.5	0.194895102691381\\
				0.51	0.194178153775576\\
				0.52	0.193463503290464\\
				0.53	0.192752222070806\\
				0.54	0.192045236115408\\
				0.55	0.191343343341375\\
				0.56	0.190647228406545\\
				0.57	0.189957475827431\\
				0.58	0.189274581592897\\
				0.59	0.188598963449918\\
				0.6	0.187930970016672\\
				0.61	0.187270888859675\\
				0.62	0.186618953655346\\
				0.63	0.185975350541998\\
				0.64	0.185340223755682\\
				0.65	0.184713680632167\\
				0.66	0.184095796047635\\
				0.67	0.183486616362078\\
				0.68	0.182886162921858\\
				0.69	0.182294435171307\\
				0.7	0.181711413417372\\
				0.71	0.181137061286249\\
				0.72	0.180571327906394\\
				0.73	0.180014149848386\\
				0.74	0.179465452848574\\
				0.75	0.178925153340411\\
				0.76	0.178393159814622\\
				0.77	0.177869374026995\\
				0.78	0.17735369207044\\
				0.79	0.176846005326126\\
				0.8	0.176346201306823\\
				0.81	0.175854164404138\\
				0.82	0.17536977655005\\
				0.83	0.174892917801976\\
				0.84	0.174423466859616\\
				0.85	0.173961301520913\\
				0.86	0.17350629908367\\
				0.87	0.17305833669865\\
				0.88	0.172617291679387\\
				0.89	0.172183041773321\\
				0.9	0.171755465398444\\
				0.91	0.171334441849135\\
				0.92	0.170919851474515\\
				0.93	0.170511575832285\\
				0.94	0.170109497820687\\
				0.95	0.169713501790973\\
				0.96	0.1693234736425\\
				0.97	0.168939300902342\\
				0.98	0.168560872791146\\
				0.99	0.168188080276723\\
				1	0.167820816116773\\
			};

		\addplot [color=mycolor4, line width=1pt]
		table[row sep=crcr]{%
				-1	-0.26216517657644\\
				-0.99	-0.262360810704833\\
				-0.98	-0.262556105203466\\
				-0.97	-0.262750930418822\\
				-0.96	-0.262945147268932\\
				-0.95	-0.263138606596774\\
				-0.94	-0.263331148477141\\
				-0.93	-0.263522601473424\\
				-0.92	-0.263712781840448\\
				-0.91	-0.263901492669187\\
				-0.9	-0.264088522968857\\
				-0.89	-0.264273646681501\\
				-0.88	-0.26445662162379\\
				-0.87	-0.264637188350335\\
				-0.86	-0.26481506893231\\
				-0.85	-0.264989965644724\\
				-0.84	-0.265161559555071\\
				-0.83	-0.265329509005542\\
				-0.82	-0.265493447980306\\
				-0.81	-0.265652984348673\\
				-0.8	-0.265807697974204\\
				-0.79	-0.265957138679008\\
				-0.78	-0.266100824051586\\
				-0.77	-0.266238237085635\\
				-0.76	-0.266368823636181\\
				-0.75	-0.266491989678341\\
				-0.74	-0.266607098352772\\
				-0.73	-0.266713466780663\\
				-0.72	-0.266810362629693\\
				-0.71	-0.266897000410959\\
				-0.7	-0.266972537485328\\
				-0.69	-0.267036069755998\\
				-0.68	-0.267086627022306\\
				-0.67	-0.267123167968007\\
				-0.66	-0.267144574755288\\
				-0.65	-0.267149647193774\\
				-0.64	-0.267137096451715\\
				-0.63	-0.267105538274383\\
				-0.62	-0.267053485672566\\
				-0.61	-0.266979341041843\\
				-0.6	-0.266881387671241\\
				-0.59	-0.266757780597802\\
				-0.58	-0.266606536761738\\
				-0.57	-0.266425524415206\\
				-0.56	-0.266212451736454\\
				-0.55	-0.26596485460029\\
				-0.54	-0.265680083455623\\
				-0.53	-0.265355289261534\\
				-0.52	-0.264987408435029\\
				-0.51	-0.26457314676673\\
				-0.5	-0.26410896226558\\
				-0.49	-0.263591046900546\\
				-0.48	-0.263015307216887\\
				-0.47	-0.262377343817301\\
				-0.46	-0.261672429714976\\
				-0.45	-0.26089548758696\\
				-0.44	-0.260041065983357\\
				-0.43	-0.259103314581674\\
				-0.42	-0.258075958617473\\
				-0.41	-0.256952272673679\\
				-0.4	-0.255725054073022\\
				-0.39	-0.254386596192853\\
				-0.38	-0.252928662110704\\
				-0.37	-0.251342459094421\\
				-0.36	-0.249618614574263\\
				-0.35	-0.247747154377751\\
				-0.34	-0.2457174841728\\
				-0.33	-0.243518375251585\\
				-0.32	-0.24113795599691\\
				-0.31	-0.238563710603499\\
				-0.3	-0.235782486875911\\
				-0.29	-0.232780515187847\\
				-0.28	-0.229543440956598\\
				-0.27	-0.226056373249832\\
				-0.26	-0.222303952383876\\
				-0.25	-0.218270439571724\\
				-0.24	-0.213939831807933\\
				-0.23	-0.209296005202344\\
				-0.22	-0.204322889854784\\
				-0.21	-0.19900467905213\\
				-0.2	-0.193326075016667\\
				-0.19	-0.187272572588707\\
				-0.18	-0.180830781038406\\
				-0.17	-0.173988782633119\\
				-0.16	-0.1667365246174\\
				-0.15	-0.159066238901559\\
				-0.14	-0.150972881049733\\
				-0.13	-0.142454577208894\\
				-0.12	-0.133513064583813\\
				-0.11	-0.124154108160602\\
				-0.1	-0.114387873892318\\
				-0.09	-0.104229236808044\\
				-0.08	-0.0936980018333325\\
				-0.07	-0.0828190158356992\\
				-0.0599999999999999	-0.0716221517851234\\
				-0.0499999999999999	-0.0601421500756295\\
				-0.04	-0.0484183079489319\\
				-0.03	-0.0364940153501062\\
				-0.02	-0.0244161439714482\\
				-0.01	-0.0122343050595997\\
				0	0\\
				0.01	0.0122343050595997\\
				0.02	0.0244161439714482\\
				0.03	0.0364940153501062\\
				0.04	0.0484183079489319\\
				0.0499999999999999	0.0601421500756295\\
				0.0599999999999999	0.0716221517851234\\
				0.07	0.0828190158356992\\
				0.08	0.0936980018333325\\
				0.09	0.104229236808044\\
				0.1	0.114387873892318\\
				0.11	0.124154108160602\\
				0.12	0.133513064583813\\
				0.13	0.142454577208894\\
				0.14	0.150972881049733\\
				0.15	0.159066238901559\\
				0.16	0.1667365246174\\
				0.17	0.173988782633119\\
				0.18	0.180830781038406\\
				0.19	0.187272572588707\\
				0.2	0.193326075016667\\
				0.21	0.19900467905213\\
				0.22	0.204322889854784\\
				0.23	0.209296005202344\\
				0.24	0.213939831807933\\
				0.25	0.218270439571724\\
				0.26	0.222303952383876\\
				0.27	0.226056373249832\\
				0.28	0.229543440956598\\
				0.29	0.232780515187847\\
				0.3	0.235782486875911\\
				0.31	0.238563710603499\\
				0.32	0.24113795599691\\
				0.33	0.243518375251585\\
				0.34	0.2457174841728\\
				0.35	0.247747154377751\\
				0.36	0.249618614574263\\
				0.37	0.251342459094421\\
				0.38	0.252928662110704\\
				0.39	0.254386596192853\\
				0.4	0.255725054073022\\
				0.41	0.256952272673679\\
				0.42	0.258075958617473\\
				0.43	0.259103314581674\\
				0.44	0.260041065983357\\
				0.45	0.26089548758696\\
				0.46	0.261672429714976\\
				0.47	0.262377343817301\\
				0.48	0.263015307216887\\
				0.49	0.263591046900546\\
				0.5	0.26410896226558\\
				0.51	0.26457314676673\\
				0.52	0.264987408435029\\
				0.53	0.265355289261534\\
				0.54	0.265680083455623\\
				0.55	0.26596485460029\\
				0.56	0.266212451736454\\
				0.57	0.266425524415206\\
				0.58	0.266606536761738\\
				0.59	0.266757780597802\\
				0.6	0.266881387671241\\
				0.61	0.266979341041843\\
				0.62	0.267053485672566\\
				0.63	0.267105538274383\\
				0.64	0.267137096451715\\
				0.65	0.267149647193774\\
				0.66	0.267144574755288\\
				0.67	0.267123167968007\\
				0.68	0.267086627022306\\
				0.69	0.267036069755998\\
				0.7	0.266972537485328\\
				0.71	0.266897000410959\\
				0.72	0.266810362629693\\
				0.73	0.266713466780663\\
				0.74	0.266607098352772\\
				0.75	0.266491989678341\\
				0.76	0.266368823636181\\
				0.77	0.266238237085635\\
				0.78	0.266100824051586\\
				0.79	0.265957138679008\\
				0.8	0.265807697974204\\
				0.81	0.265652984348673\\
				0.82	0.265493447980306\\
				0.83	0.265329509005542\\
				0.84	0.265161559555071\\
				0.85	0.264989965644724\\
				0.86	0.26481506893231\\
				0.87	0.264637188350335\\
				0.88	0.26445662162379\\
				0.89	0.264273646681501\\
				0.9	0.264088522968857\\
				0.91	0.263901492669187\\
				0.92	0.263712781840448\\
				0.93	0.263522601473424\\
				0.94	0.263331148477141\\
				0.95	0.263138606596774\\
				0.96	0.262945147268932\\
				0.97	0.262750930418822\\
				0.98	0.262556105203466\\
				0.99	0.262360810704833\\
				1	0.26216517657644\\
			};

		\addplot [color=mycolor5, line width=1pt]
		table[row sep=crcr]{%
				-1	-0.275533060874427\\
				-0.99	-0.276143694071453\\
				-0.98	-0.276760131305061\\
				-0.97	-0.277382354032618\\
				-0.96	-0.278010335633523\\
				-0.95	-0.278644040676319\\
				-0.94	-0.279283424125511\\
				-0.93	-0.27992843048302\\
				-0.92	-0.280578992858816\\
				-0.91	-0.281235031964735\\
				-0.9	-0.281896455025008\\
				-0.89	-0.282563154596434\\
				-0.88	-0.283235007290539\\
				-0.87	-0.283911872389346\\
				-0.86	-0.28459359034568\\
				-0.85	-0.28527998115812\\
				-0.84	-0.285970842609818\\
				-0.83	-0.286665948359482\\
				-0.82	-0.287365045871755\\
				-0.81	-0.288067854173112\\
				-0.8	-0.288774061418149\\
				-0.79	-0.289483322249818\\
				-0.78	-0.290195254935697\\
				-0.77	-0.290909438260799\\
				-0.76	-0.291625408155723\\
				-0.75	-0.292342654037071\\
				-0.74	-0.293060614835054\\
				-0.73	-0.293778674681008\\
				-0.72	-0.294496158225203\\
				-0.71	-0.295212325552775\\
				-0.7	-0.295926366662852\\
				-0.69	-0.296637395473037\\
				-0.68	-0.297344443308225\\
				-0.67	-0.29804645182937\\
				-0.66	-0.298742265354245\\
				-0.65	-0.299430622518407\\
				-0.64	-0.300110147220585\\
				-0.63	-0.300779338792517\\
				-0.62	-0.301436561328846\\
				-0.61	-0.302080032108205\\
				-0.6	-0.302707809032002\\
				-0.59	-0.303317777002775\\
				-0.58	-0.303907633159411\\
				-0.57	-0.304474870882133\\
				-0.56	-0.305016762476029\\
				-0.55	-0.305530340438346\\
				-0.54	-0.30601237721184\\
				-0.53	-0.306459363324676\\
				-0.52	-0.306867483816838\\
				-0.51	-0.307232592854352\\
				-0.5	-0.307550186436312\\
				-0.49	-0.307815373106362\\
				-0.48	-0.30802284259077\\
				-0.47	-0.308166832300487\\
				-0.46	-0.30824109165565\\
				-0.45	-0.308238844219372\\
				-0.44	-0.30815274766479\\
				-0.43	-0.307974851647145\\
				-0.42	-0.30769655371326\\
				-0.41	-0.307308553456568\\
				-0.4	-0.306800805219616\\
				-0.39	-0.306162469760731\\
				-0.38	-0.305381865440657\\
				-0.37	-0.304446419651947\\
				-0.36	-0.303342621412428\\
				-0.35	-0.302055976277634\\
				-0.34	-0.300570964999069\\
				-0.33	-0.298871007668087\\
				-0.32	-0.296938435440415\\
				-0.31	-0.294754472333722\\
				-0.3	-0.292299230027068\\
				-0.29	-0.289551719060079\\
				-0.28	-0.28648988031997\\
				-0.27	-0.283090641198462\\
				-0.26	-0.279330001272889\\
				-0.25	-0.275183152781666\\
				-0.24	-0.270624641478245\\
				-0.23	-0.265628573602072\\
				-0.22	-0.260168874629971\\
				-0.21	-0.254219605084985\\
				-0.2	-0.247755337891086\\
				-0.19	-0.240751600475834\\
				-0.18	-0.233185382946758\\
				-0.17	-0.225035711123552\\
				-0.16	-0.216284279948773\\
				-0.15	-0.206916138824544\\
				-0.14	-0.196920415799488\\
				-0.13	-0.186291062415282\\
				-0.12	-0.175027595677605\\
				-0.11	-0.163135808417602\\
				-0.1	-0.150628414742641\\
				-0.09	-0.137525593910359\\
				-0.08	-0.123855394410041\\
				-0.07	-0.109653960884792\\
				-0.0599999999999999	-0.0949655502473018\\
				-0.0499999999999999	-0.0798423101910752\\
				-0.04	-0.0643438032385393\\
				-0.03	-0.0485362720939867\\
				-0.02	-0.0324916565973149\\
				-0.01	-0.0162863878757812\\
				0	0\\
				0.01	0.0162863878757812\\
				0.02	0.0324916565973149\\
				0.03	0.0485362720939867\\
				0.04	0.0643438032385393\\
				0.0499999999999999	0.0798423101910752\\
				0.0599999999999999	0.0949655502473018\\
				0.07	0.109653960884792\\
				0.08	0.123855394410041\\
				0.09	0.137525593910359\\
				0.1	0.150628414742641\\
				0.11	0.163135808417602\\
				0.12	0.175027595677605\\
				0.13	0.186291062415282\\
				0.14	0.196920415799488\\
				0.15	0.206916138824544\\
				0.16	0.216284279948773\\
				0.17	0.225035711123552\\
				0.18	0.233185382946758\\
				0.19	0.240751600475834\\
				0.2	0.247755337891086\\
				0.21	0.254219605084985\\
				0.22	0.260168874629971\\
				0.23	0.265628573602072\\
				0.24	0.270624641478245\\
				0.25	0.275183152781666\\
				0.26	0.279330001272889\\
				0.27	0.283090641198462\\
				0.28	0.28648988031997\\
				0.29	0.289551719060079\\
				0.3	0.292299230027068\\
				0.31	0.294754472333722\\
				0.32	0.296938435440415\\
				0.33	0.298871007668087\\
				0.34	0.300570964999069\\
				0.35	0.302055976277634\\
				0.36	0.303342621412428\\
				0.37	0.304446419651947\\
				0.38	0.305381865440657\\
				0.39	0.306162469760731\\
				0.4	0.306800805219616\\
				0.41	0.307308553456568\\
				0.42	0.30769655371326\\
				0.43	0.307974851647145\\
				0.44	0.30815274766479\\
				0.45	0.308238844219372\\
				0.46	0.30824109165565\\
				0.47	0.308166832300487\\
				0.48	0.30802284259077\\
				0.49	0.307815373106362\\
				0.5	0.307550186436312\\
				0.51	0.307232592854352\\
				0.52	0.306867483816838\\
				0.53	0.306459363324676\\
				0.54	0.30601237721184\\
				0.55	0.305530340438346\\
				0.56	0.305016762476029\\
				0.57	0.304474870882133\\
				0.58	0.303907633159411\\
				0.59	0.303317777002775\\
				0.6	0.302707809032002\\
				0.61	0.302080032108205\\
				0.62	0.301436561328846\\
				0.63	0.300779338792517\\
				0.64	0.300110147220585\\
				0.65	0.299430622518407\\
				0.66	0.298742265354245\\
				0.67	0.29804645182937\\
				0.68	0.297344443308225\\
				0.69	0.296637395473037\\
				0.7	0.295926366662852\\
				0.71	0.295212325552775\\
				0.72	0.294496158225203\\
				0.73	0.293778674681008\\
				0.74	0.293060614835054\\
				0.75	0.292342654037071\\
				0.76	0.291625408155723\\
				0.77	0.290909438260799\\
				0.78	0.290195254935697\\
				0.79	0.289483322249818\\
				0.8	0.288774061418149\\
				0.81	0.288067854173112\\
				0.82	0.287365045871755\\
				0.83	0.286665948359482\\
				0.84	0.285970842609818\\
				0.85	0.28527998115812\\
				0.86	0.28459359034568\\
				0.87	0.283911872389346\\
				0.88	0.283235007290539\\
				0.89	0.282563154596434\\
				0.9	0.281896455025008\\
				0.91	0.281235031964735\\
				0.92	0.280578992858816\\
				0.93	0.27992843048302\\
				0.94	0.279283424125511\\
				0.95	0.278644040676319\\
				0.96	0.278010335633523\\
				0.97	0.277382354032618\\
				0.98	0.276760131305061\\
				0.99	0.276143694071453\\
				1	0.275533060874427\\
			};

		\addplot [color=mycolor6, line width=1pt]
		table[row sep=crcr]{%
				-1	-0.307888835002677\\
				-0.99	-0.307842658546626\\
				-0.98	-0.307792752829448\\
				-0.97	-0.307738929439374\\
				-0.96	-0.307680989765714\\
				-0.95	-0.307618724380415\\
				-0.94	-0.307551912377777\\
				-0.93	-0.307480320669176\\
				-0.92	-0.307403703229458\\
				-0.91	-0.307321800291325\\
				-0.9	-0.307234337483804\\
				-0.89	-0.307141024910504\\
				-0.88	-0.307041556163069\\
				-0.87	-0.306935607264823\\
				-0.86	-0.306822835539198\\
				-0.85	-0.306702878397093\\
				-0.84	-0.306575352036805\\
				-0.83	-0.306439850049678\\
				-0.82	-0.30629594192399\\
				-0.81	-0.306143171439022\\
				-0.8	-0.305981054940543\\
				-0.79	-0.305809079488202\\
				-0.78	-0.305626700864542\\
				-0.77	-0.305433341434472\\
				-0.76	-0.305228387843057\\
				-0.75	-0.305011188538524\\
				-0.74	-0.304781051106215\\
				-0.73	-0.304537239398071\\
				-0.72	-0.304278970440883\\
				-0.71	-0.304005411105201\\
				-0.7	-0.303715674515241\\
				-0.69	-0.303408816178547\\
				-0.68	-0.303083829812378\\
				-0.67	-0.302739642841971\\
				-0.66	-0.302375111543808\\
				-0.65	-0.301989015804905\\
				-0.64	-0.301580053466878\\
				-0.63	-0.301146834221188\\
				-0.62	-0.300687873019465\\
				-0.61	-0.300201582960187\\
				-0.6	-0.299686267610358\\
				-0.59	-0.299140112718049\\
				-0.58	-0.298561177268887\\
				-0.57	-0.297947383836899\\
				-0.56	-0.297296508177381\\
				-0.55	-0.296606168007083\\
				-0.54	-0.295873810914737\\
				-0.53	-0.295096701343251\\
				-0.52	-0.294271906583702\\
				-0.51	-0.293396281720937\\
				-0.5	-0.292466453471331\\
				-0.49	-0.291478802855458\\
				-0.48	-0.290429446652428\\
				-0.47	-0.289314217589078\\
				-0.46	-0.28812864322651\\
				-0.45	-0.286867923519598\\
				-0.44	-0.285526907042739\\
				-0.43	-0.284100065898559\\
				-0.42	-0.28258146935658\\
				-0.41	-0.280964756307707\\
				-0.4	-0.279243106669394\\
				-0.39	-0.277409211937557\\
				-0.38	-0.275455245157073\\
				-0.37	-0.273372830675542\\
				-0.36	-0.27115301415772\\
				-0.35	-0.268786233473825\\
				-0.34	-0.266262291236814\\
				-0.33	-0.263570329954923\\
				-0.32	-0.260698810989231\\
				-0.31	-0.257635498763856\\
				-0.3	-0.254367451970099\\
				-0.29	-0.250881023834786\\
				-0.28	-0.247161873884375\\
				-0.27	-0.243194994023595\\
				-0.26	-0.23896475214908\\
				-0.25	-0.234454956917026\\
				-0.24	-0.229648947653604\\
				-0.23	-0.224529713702207\\
				-0.22	-0.219080047695724\\
				-0.21	-0.213282737265404\\
				-0.2	-0.207120799478215\\
				-0.19	-0.200577761747931\\
				-0.18	-0.193637991999493\\
				-0.17	-0.186287079387872\\
				-0.16	-0.178512264797263\\
				-0.15	-0.170302917613446\\
				-0.14	-0.161651051854511\\
				-0.13	-0.152551870712353\\
				-0.12	-0.143004324039325\\
				-0.11	-0.133011658562885\\
				-0.1	-0.122581936002034\\
				-0.09	-0.111728490289735\\
				-0.08	-0.100470292367213\\
				-0.07	-0.0888321901411228\\
				-0.0599999999999999	-0.0768449927720445\\
				-0.0499999999999999	-0.0645453729314488\\
				-0.04	-0.0519755681982976\\
				-0.03	-0.0391828731743503\\
				-0.02	-0.0262189265634243\\
				-0.01	-0.0131388113529625\\
				0	0\\
				0.01	0.0131388113529625\\
				0.02	0.0262189265634243\\
				0.03	0.0391828731743503\\
				0.04	0.0519755681982976\\
				0.0499999999999999	0.0645453729314488\\
				0.0599999999999999	0.0768449927720445\\
				0.07	0.0888321901411228\\
				0.08	0.100470292367213\\
				0.09	0.111728490289735\\
				0.1	0.122581936002034\\
				0.11	0.133011658562885\\
				0.12	0.143004324039325\\
				0.13	0.152551870712353\\
				0.14	0.161651051854511\\
				0.15	0.170302917613446\\
				0.16	0.178512264797263\\
				0.17	0.186287079387872\\
				0.18	0.193637991999493\\
				0.19	0.200577761747931\\
				0.2	0.207120799478215\\
				0.21	0.213282737265404\\
				0.22	0.219080047695724\\
				0.23	0.224529713702207\\
				0.24	0.229648947653604\\
				0.25	0.234454956917026\\
				0.26	0.23896475214908\\
				0.27	0.243194994023595\\
				0.28	0.247161873884375\\
				0.29	0.250881023834786\\
				0.3	0.254367451970099\\
				0.31	0.257635498763856\\
				0.32	0.260698810989231\\
				0.33	0.263570329954923\\
				0.34	0.266262291236814\\
				0.35	0.268786233473825\\
				0.36	0.27115301415772\\
				0.37	0.273372830675542\\
				0.38	0.275455245157073\\
				0.39	0.277409211937557\\
				0.4	0.279243106669394\\
				0.41	0.280964756307707\\
				0.42	0.28258146935658\\
				0.43	0.284100065898559\\
				0.44	0.285526907042739\\
				0.45	0.286867923519598\\
				0.46	0.28812864322651\\
				0.47	0.289314217589078\\
				0.48	0.290429446652428\\
				0.49	0.291478802855458\\
				0.5	0.292466453471331\\
				0.51	0.293396281720937\\
				0.52	0.294271906583702\\
				0.53	0.295096701343251\\
				0.54	0.295873810914737\\
				0.55	0.296606168007083\\
				0.56	0.297296508177381\\
				0.57	0.297947383836899\\
				0.58	0.298561177268887\\
				0.59	0.299140112718049\\
				0.6	0.299686267610358\\
				0.61	0.300201582960187\\
				0.62	0.300687873019465\\
				0.63	0.301146834221188\\
				0.64	0.301580053466878\\
				0.65	0.301989015804905\\
				0.66	0.302375111543808\\
				0.67	0.302739642841971\\
				0.68	0.303083829812378\\
				0.69	0.303408816178547\\
				0.7	0.303715674515241\\
				0.71	0.304005411105201\\
				0.72	0.304278970440883\\
				0.73	0.304537239398071\\
				0.74	0.304781051106215\\
				0.75	0.305011188538524\\
				0.76	0.305228387843057\\
				0.77	0.305433341434472\\
				0.78	0.305626700864542\\
				0.79	0.305809079488202\\
				0.8	0.305981054940543\\
				0.81	0.306143171439022\\
				0.82	0.30629594192399\\
				0.83	0.306439850049678\\
				0.84	0.306575352036805\\
				0.85	0.306702878397093\\
				0.86	0.306822835539198\\
				0.87	0.306935607264823\\
				0.88	0.307041556163069\\
				0.89	0.307141024910504\\
				0.9	0.307234337483804\\
				0.91	0.307321800291325\\
				0.92	0.307403703229458\\
				0.93	0.307480320669176\\
				0.94	0.307551912377777\\
				0.95	0.307618724380415\\
				0.96	0.307680989765714\\
				0.97	0.307738929439374\\
				0.98	0.307792752829448\\
				0.99	0.307842658546626\\
				1	0.307888835002677\\
			};

		\addplot [color=mycolor7, line width=1pt]
		table[row sep=crcr]{%
				-1	-0.259810324268026\\
				-0.99	-0.260395098654399\\
				-0.98	-0.260987832508282\\
				-0.97	-0.261588645793898\\
				-0.96	-0.262197658333836\\
				-0.95	-0.262814989550274\\
				-0.94	-0.263440758175624\\
				-0.93	-0.264075081929439\\
				-0.92	-0.264718077158091\\
				-0.91	-0.265369858433357\\
				-0.9	-0.26603053810566\\
				-0.89	-0.266700225807241\\
				-0.88	-0.267379027900072\\
				-0.87	-0.268067046862722\\
				-0.86	-0.268764380609815\\
				-0.85	-0.269471121737004\\
				-0.84	-0.270187356683646\\
				-0.83	-0.270913164804491\\
				-0.82	-0.271648617340788\\
				-0.81	-0.272393776280124\\
				-0.8	-0.273148693093171\\
				-0.79	-0.273913407334203\\
				-0.78	-0.274687945090776\\
				-0.77	-0.275472317266363\\
				-0.76	-0.276266517677908\\
				-0.75	-0.27707052094825\\
				-0.74	-0.277884280171097\\
				-0.73	-0.278707724323726\\
				-0.72	-0.279540755399732\\
				-0.71	-0.280383245231025\\
				-0.7	-0.281235031964735\\
				-0.69	-0.282095916156727\\
				-0.68	-0.282965656439037\\
				-0.67	-0.283843964713596\\
				-0.66	-0.284730500819075\\
				-0.65	-0.285624866611539\\
				-0.64	-0.286526599392657\\
				-0.63	-0.287435164611523\\
				-0.62	-0.288349947757514\\
				-0.61	-0.289270245351963\\
				-0.6	-0.290195254935697\\
				-0.59	-0.291124063937491\\
				-0.58	-0.292055637295154\\
				-0.57	-0.29298880368611\\
				-0.56	-0.293922240207886\\
				-0.55	-0.294854455330635\\
				-0.54	-0.295783769923705\\
				-0.53	-0.296708296136008\\
				-0.52	-0.29762591388562\\
				-0.51	-0.298534244687383\\
				-0.5	-0.299430622518407\\
				-0.49	-0.300312061390162\\
				-0.48	-0.301175219262522\\
				-0.47	-0.302016357899862\\
				-0.46	-0.302831298232572\\
				-0.45	-0.303615370749756\\
				-0.44	-0.304363360411458\\
				-0.43	-0.305069445532879\\
				-0.42	-0.305727130060684\\
				-0.41	-0.306329168635312\\
				-0.4	-0.306867483816838\\
				-0.39	-0.307333074850079\\
				-0.38	-0.307715917363673\\
				-0.37	-0.308004853445765\\
				-0.36	-0.308187471626219\\
				-0.35	-0.308249976435116\\
				-0.34	-0.308177047416239\\
				-0.33	-0.307951687772855\\
				-0.32	-0.307555063236561\\
				-0.31	-0.306966332308403\\
				-0.3	-0.306162469760731\\
				-0.29	-0.305118086249332\\
				-0.28	-0.303805248114394\\
				-0.27	-0.302193302994634\\
				-0.26	-0.300248718790628\\
				-0.25	-0.297934945834475\\
				-0.24	-0.295212314883009\\
				-0.23	-0.292037986754385\\
				-0.22	-0.288365973032716\\
				-0.21	-0.284147251165656\\
				-0.2	-0.279330001272889\\
				-0.19	-0.273859995736867\\
				-0.18	-0.267681175662576\\
				-0.17	-0.260736449874072\\
				-0.16	-0.252968751349529\\
				-0.15	-0.244322381766573\\
				-0.14	-0.234744665876663\\
				-0.13	-0.224187922485368\\
				-0.12	-0.212611736844099\\
				-0.11	-0.199985489779956\\
				-0.1	-0.186291062415282\\
				-0.09	-0.171525593818891\\
				-0.08	-0.155704126179166\\
				-0.07	-0.138861933853899\\
				-0.0599999999999999	-0.121056306351716\\
				-0.0499999999999999	-0.102367549104411\\
				-0.04	-0.0828989871622466\\
				-0.03	-0.0627758102182411\\
				-0.02	-0.0421426822937078\\
				-0.01	-0.0211601493498722\\
				0	0\\
				0.01	0.0211601493498722\\
				0.02	0.0421426822937078\\
				0.03	0.0627758102182411\\
				0.04	0.0828989871622466\\
				0.0499999999999999	0.102367549104411\\
				0.0599999999999999	0.121056306351716\\
				0.07	0.138861933853899\\
				0.08	0.155704126179166\\
				0.09	0.171525593818891\\
				0.1	0.186291062415282\\
				0.11	0.199985489779956\\
				0.12	0.212611736844099\\
				0.13	0.224187922485368\\
				0.14	0.234744665876663\\
				0.15	0.244322381766573\\
				0.16	0.252968751349529\\
				0.17	0.260736449874072\\
				0.18	0.267681175662576\\
				0.19	0.273859995736867\\
				0.2	0.279330001272889\\
				0.21	0.284147251165656\\
				0.22	0.288365973032716\\
				0.23	0.292037986754385\\
				0.24	0.295212314883009\\
				0.25	0.297934945834475\\
				0.26	0.300248718790628\\
				0.27	0.302193302994634\\
				0.28	0.303805248114394\\
				0.29	0.305118086249332\\
				0.3	0.306162469760731\\
				0.31	0.306966332308403\\
				0.32	0.307555063236561\\
				0.33	0.307951687772855\\
				0.34	0.308177047416239\\
				0.35	0.308249976435116\\
				0.36	0.308187471626219\\
				0.37	0.308004853445765\\
				0.38	0.307715917363673\\
				0.39	0.307333074850079\\
				0.4	0.306867483816838\\
				0.41	0.306329168635312\\
				0.42	0.305727130060684\\
				0.43	0.305069445532879\\
				0.44	0.304363360411458\\
				0.45	0.303615370749756\\
				0.46	0.302831298232572\\
				0.47	0.302016357899862\\
				0.48	0.301175219262522\\
				0.49	0.300312061390162\\
				0.5	0.299430622518407\\
				0.51	0.298534244687383\\
				0.52	0.29762591388562\\
				0.53	0.296708296136008\\
				0.54	0.295783769923705\\
				0.55	0.294854455330635\\
				0.56	0.293922240207886\\
				0.57	0.29298880368611\\
				0.58	0.292055637295154\\
				0.59	0.291124063937491\\
				0.6	0.290195254935697\\
				0.61	0.289270245351963\\
				0.62	0.288349947757514\\
				0.63	0.287435164611523\\
				0.64	0.286526599392657\\
				0.65	0.285624866611539\\
				0.66	0.284730500819075\\
				0.67	0.283843964713596\\
				0.68	0.282965656439037\\
				0.69	0.282095916156727\\
				0.7	0.281235031964735\\
				0.71	0.280383245231025\\
				0.72	0.279540755399732\\
				0.73	0.278707724323726\\
				0.74	0.277884280171097\\
				0.75	0.27707052094825\\
				0.76	0.276266517677908\\
				0.77	0.275472317266363\\
				0.78	0.274687945090776\\
				0.79	0.273913407334203\\
				0.8	0.273148693093171\\
				0.81	0.272393776280124\\
				0.82	0.271648617340788\\
				0.83	0.270913164804491\\
				0.84	0.270187356683646\\
				0.85	0.269471121737004\\
				0.86	0.268764380609815\\
				0.87	0.268067046862722\\
				0.88	0.267379027900072\\
				0.89	0.266700225807241\\
				0.9	0.26603053810566\\
				0.91	0.265369858433357\\
				0.92	0.264718077158091\\
				0.93	0.264075081929439\\
				0.94	0.263440758175624\\
				0.95	0.262814989550274\\
				0.96	0.262197658333836\\
				0.97	0.261588645793898\\
				0.98	0.260987832508282\\
				0.99	0.260395098654399\\
				1	0.259810324268026\\
			};

	\end{axis}
\end{tikzpicture}%

%% file: car_lap_alpaca.tex
% This file was created by tikzplotlib v0.9.1.
\begin{tikzpicture}

	\definecolor{color0}{rgb}{0.12156862745098,0.466666666666667,0.705882352941177}

	\begin{axis}[
			legend cell align={left},
			legend columns = -1,
			legend style={fill opacity=0.8, draw opacity=1, text opacity=1, at={(0.03,0.97)}, draw=white!80!black, at={(1.17,1.25)}},
			tick align=outside,
			tick pos=left,
			x grid style={white!69.0196078431373!black},
			xmin=-1.903, xmax=1.903,
			xtick style={color=black},
			y grid style={white!69.0196078431373!black},
			ymin=-1.408, ymax=1.408,
			ytick style={color=black},
			scale={0.45},
		]
		\addplot [semithick, black, forget plot]
		table {%
				0.149999976158142 -1.27999997138977
				1.22377836704254 -1.27946627140045
				1.25934112071991 -1.27666747570038
				1.29463517665863 -1.27148270606995
				1.32950043678284 -1.2639354467392
				1.36377894878387 -1.25405991077423
				1.39731562137604 -1.24190092086792
				1.42995834350586 -1.22751355171204
				1.46155941486359 -1.21096277236938
				1.49197542667389 -1.19232380390167
				1.52106869220734 -1.1716810464859
				1.54870748519897 -1.14912784099579
				1.57476663589478 -1.12476658821106
				1.5991278886795 -1.09870755672455
				1.62168097496033 -1.07106876373291
				1.64232385158539 -1.04197537899017
				1.6609628200531 -1.01155936717987
				1.67751348018646 -0.979958295822144
				1.69190096855164 -0.947315692901611
				1.70405995845795 -0.913779020309448
				1.71393549442291 -0.879500389099121
				1.72148263454437 -0.844635128974915
				1.72666752338409 -0.809341192245483
				1.72946631908417 -0.773778438568115
				1.73000001907349 -0.730000019073486
				1.72946631908417 0.773778438568115
				1.72666752338409 0.809341192245483
				1.72148263454437 0.844635128974915
				1.71393549442291 0.879500389099121
				1.70405995845795 0.913779020309448
				1.69190096855164 0.947315692901611
				1.67751348018646 0.979958295822144
				1.6609628200531 1.01155936717987
				1.64232385158539 1.04197537899017
				1.62168097496033 1.07106876373291
				1.5991278886795 1.09870755672455
				1.57476663589478 1.12476658821106
				1.54870748519897 1.14912784099579
				1.52106869220734 1.1716810464859
				1.49197542667389 1.19232380390167
				1.46155941486359 1.21096277236938
				1.42995834350586 1.22751355171204
				1.39731562137604 1.24190092086792
				1.36377894878387 1.25405991077423
				1.32950043678284 1.2639354467392
				1.29463517665863 1.27148270606995
				1.25934112071991 1.27666747570038
				1.22377836704254 1.27946627140045
				1.17999994754791 1.27999997138977
				0.876221656799316 1.27946627140045
				0.840658903121948 1.27666747570038
				0.805364847183228 1.27148270606995
				0.770499587059021 1.2639354467392
				0.736221075057983 1.25405991077423
				0.70268440246582 1.24190092086792
				0.670041561126709 1.22751355171204
				0.638440608978271 1.21096277236938
				0.608024597167969 1.19232380390167
				0.578931331634521 1.1716810464859
				0.551292419433594 1.14912784099579
				0.525233387947083 1.12476658821106
				0.500872135162354 1.09870755672455
				0.478318929672241 1.07106876373291
				0.45767617225647 1.04197537899017
				0.439037203788757 1.01155936717987
				0.422486543655396 0.979958295822144
				0.408099055290222 0.947315692901611
				0.395940065383911 0.913779020309448
				0.386064529418945 0.879500389099121
				0.378517389297485 0.844635128974915
				0.373332500457764 0.809341192245483
				0.37053370475769 0.773778438568115
				0.370000004768372 0.730000019073486
				0.369138240814209 0.139049530029297
				0.366573929786682 0.128368854522705
				0.362370491027832 0.118220686912537
				0.35663115978241 0.108855009078979
				0.349497437477112 0.100502490997314
				0.341145038604736 0.0933687686920166
				0.331779360771179 0.0876295566558838
				0.321631193161011 0.0834259986877441
				0.310950398445129 0.0808618068695068
				0.299999952316284 0.0800000429153442
				-0.0109504461288452 0.0808618068695068
				-0.0216312408447266 0.0834259986877441
				-0.0317792892456055 0.0876295566558838
				-0.0411449670791626 0.0933687686920166
				-0.0494974851608276 0.100502490997314
				-0.0566312074661255 0.108855009078979
				-0.0623704195022583 0.118220686912537
				-0.0665739774703979 0.128368854522705
				-0.0691381692886353 0.139049530029297
				-0.0700000524520874 0.149999976158142
				-0.0705336332321167 0.773778438568115
				-0.0733325481414795 0.809341192245483
				-0.0785173177719116 0.844635128974915
				-0.0860645771026611 0.879500389099121
				-0.0959399938583374 0.913779020309448
				-0.108099102973938 0.947315692901611
				-0.122486472129822 0.979958295822144
				-0.139037251472473 1.01155936717987
				-0.157676219940186 1.04197537899017
				-0.178318977355957 1.07106876373291
				-0.200872182846069 1.09870755672455
				-0.225233435630798 1.12476658821106
				-0.25129246711731 1.14912784099579
				-0.278931260108948 1.1716810464859
				-0.308024644851685 1.19232380390167
				-0.338440656661987 1.21096277236938
				-0.370041608810425 1.22751355171204
				-0.402684330940247 1.24190092086792
				-0.43622100353241 1.25405991077423
				-0.470499515533447 1.2639354467392
				-0.505364894866943 1.27148270606995
				-0.540658831596375 1.27666747570038
				-0.576221704483032 1.27946627140045
				-0.620224714279175 1.27999997138977
				-1.22377836704254 1.27946627140045
				-1.25934112071991 1.27666747570038
				-1.29463517665863 1.27148270606995
				-1.32950043678284 1.2639354467392
				-1.36377894878387 1.25405991077423
				-1.39731562137604 1.24190092086792
				-1.42995834350586 1.22751355171204
				-1.46155941486359 1.21096277236938
				-1.49197542667389 1.19232380390167
				-1.52106869220734 1.1716810464859
				-1.54870748519897 1.14912784099579
				-1.57476663589478 1.12476658821106
				-1.5991278886795 1.09870755672455
				-1.62168097496033 1.07106876373291
				-1.64232385158539 1.04197537899017
				-1.6609628200531 1.01155936717987
				-1.67751348018646 0.979958295822144
				-1.69190096855164 0.947315692901611
				-1.70405995845795 0.913779020309448
				-1.71393549442291 0.879500389099121
				-1.72148263454437 0.844635128974915
				-1.72666752338409 0.809341192245483
				-1.72946631908417 0.773778438568115
				-1.73000001907349 0.730000019073486
				-1.72946631908417 -0.773778438568115
				-1.72666752338409 -0.809341192245483
				-1.72148263454437 -0.844635128974915
				-1.71393549442291 -0.879500389099121
				-1.70405995845795 -0.913779020309448
				-1.69190096855164 -0.947315692901611
				-1.67751348018646 -0.979958295822144
				-1.6609628200531 -1.01155936717987
				-1.64232385158539 -1.04197537899017
				-1.62168097496033 -1.07106876373291
				-1.5991278886795 -1.09870755672455
				-1.57476663589478 -1.12476658821106
				-1.54870748519897 -1.14912784099579
				-1.52106869220734 -1.1716810464859
				-1.49197542667389 -1.19232380390167
				-1.46155941486359 -1.21096277236938
				-1.42995834350586 -1.22751355171204
				-1.39731562137604 -1.24190092086792
				-1.36377894878387 -1.25405991077423
				-1.32950043678284 -1.2639354467392
				-1.29463517665863 -1.27148270606995
				-1.25934112071991 -1.27666747570038
				-1.22377836704254 -1.27946627140045
				-1.17995047569275 -1.27999997138977
				1.22377836704254 -1.27946627140045
				1.25934112071991 -1.27666747570038
				1.29463517665863 -1.27148270606995
				1.32950043678284 -1.2639354467392
				1.36377894878387 -1.25405991077423
				1.39731562137604 -1.24190092086792
				1.42995834350586 -1.22751355171204
				1.46155941486359 -1.21096277236938
				1.49197542667389 -1.19232380390167
				1.52106869220734 -1.1716810464859
				1.54870748519897 -1.14912784099579
				1.57476663589478 -1.12476658821106
				1.5991278886795 -1.09870755672455
				1.62168097496033 -1.07106876373291
				1.64232385158539 -1.04197537899017
				1.6609628200531 -1.01155936717987
				1.67751348018646 -0.979958295822144
				1.69190096855164 -0.947315692901611
				1.70405995845795 -0.913779020309448
				1.71393549442291 -0.879500389099121
				1.72148263454437 -0.844635128974915
				1.72666752338409 -0.809341192245483
				1.72946631908417 -0.773778438568115
				1.73000001907349 -0.730000019073486
				1.72946631908417 0.773778438568115
				1.72666752338409 0.809341192245483
				1.72148263454437 0.844635128974915
				1.71393549442291 0.879500389099121
				1.70405995845795 0.913779020309448
				1.69190096855164 0.947315692901611
				1.67751348018646 0.979958295822144
				1.6609628200531 1.01155936717987
				1.64232385158539 1.04197537899017
				1.62168097496033 1.07106876373291
				1.5991278886795 1.09870755672455
				1.57476663589478 1.12476658821106
				1.54870748519897 1.14912784099579
				1.52106869220734 1.1716810464859
				1.49197542667389 1.19232380390167
				1.46155941486359 1.21096277236938
				1.42995834350586 1.22751355171204
				1.39731562137604 1.24190092086792
				1.36377894878387 1.25405991077423
				1.32950043678284 1.2639354467392
				1.29463517665863 1.27148270606995
				1.25934112071991 1.27666747570038
				1.22377836704254 1.27946627140045
				1.17999994754791 1.27999997138977
				0.876221656799316 1.27946627140045
				0.840658903121948 1.27666747570038
				0.805364847183228 1.27148270606995
				0.770499587059021 1.2639354467392
				0.736221075057983 1.25405991077423
				0.70268440246582 1.24190092086792
				0.670041561126709 1.22751355171204
				0.638440608978271 1.21096277236938
				0.608024597167969 1.19232380390167
				0.578931331634521 1.1716810464859
				0.551292419433594 1.14912784099579
				0.525233387947083 1.12476658821106
				0.500872135162354 1.09870755672455
				0.478318929672241 1.07106876373291
				0.45767617225647 1.04197537899017
				0.439037203788757 1.01155936717987
				0.422486543655396 0.979958295822144
				0.408099055290222 0.947315692901611
				0.395940065383911 0.913779020309448
				0.386064529418945 0.879500389099121
				0.378517389297485 0.844635128974915
				0.373332500457764 0.809341192245483
				0.37053370475769 0.773778438568115
				0.370000004768372 0.730000019073486
				0.369138240814209 0.139049530029297
				0.366573929786682 0.128368854522705
				0.362370491027832 0.118220686912537
				0.35663115978241 0.108855009078979
				0.349497437477112 0.100502490997314
				0.341145038604736 0.0933687686920166
				0.331779360771179 0.0876295566558838
				0.321631193161011 0.0834259986877441
				0.310950398445129 0.0808618068695068
				0.299999952316284 0.0800000429153442
				-0.0109504461288452 0.0808618068695068
				-0.0216312408447266 0.0834259986877441
				-0.0317792892456055 0.0876295566558838
				-0.0411449670791626 0.0933687686920166
				-0.0494974851608276 0.100502490997314
				-0.0566312074661255 0.108855009078979
				-0.0623704195022583 0.118220686912537
				-0.0665739774703979 0.128368854522705
				-0.0691381692886353 0.139049530029297
				-0.0700000524520874 0.149999976158142
				-0.0705336332321167 0.773778438568115
				-0.0733325481414795 0.809341192245483
				-0.0785173177719116 0.844635128974915
				-0.0860645771026611 0.879500389099121
				-0.0959399938583374 0.913779020309448
				-0.108099102973938 0.947315692901611
				-0.122486472129822 0.979958295822144
				-0.139037251472473 1.01155936717987
				-0.157676219940186 1.04197537899017
				-0.178318977355957 1.07106876373291
				-0.200872182846069 1.09870755672455
				-0.225233435630798 1.12476658821106
				-0.25129246711731 1.14912784099579
				-0.278931260108948 1.1716810464859
				-0.308024644851685 1.19232380390167
				-0.338440656661987 1.21096277236938
				-0.370041608810425 1.22751355171204
				-0.402684330940247 1.24190092086792
				-0.43622100353241 1.25405991077423
				-0.470499515533447 1.2639354467392
				-0.505364894866943 1.27148270606995
				-0.540658831596375 1.27666747570038
				-0.576221704483032 1.27946627140045
				-0.620224714279175 1.27999997138977
				-1.22377836704254 1.27946627140045
				-1.25934112071991 1.27666747570038
				-1.29463517665863 1.27148270606995
				-1.32950043678284 1.2639354467392
				-1.36377894878387 1.25405991077423
				-1.39731562137604 1.24190092086792
				-1.42995834350586 1.22751355171204
				-1.46155941486359 1.21096277236938
				-1.49197542667389 1.19232380390167
				-1.52106869220734 1.1716810464859
				-1.54870748519897 1.14912784099579
				-1.57476663589478 1.12476658821106
				-1.5991278886795 1.09870755672455
				-1.62168097496033 1.07106876373291
				-1.64232385158539 1.04197537899017
				-1.6609628200531 1.01155936717987
				-1.67751348018646 0.979958295822144
				-1.69190096855164 0.947315692901611
				-1.70405995845795 0.913779020309448
				-1.71393549442291 0.879500389099121
				-1.72148263454437 0.844635128974915
				-1.72666752338409 0.809341192245483
				-1.72946631908417 0.773778438568115
				-1.73000001907349 0.730000019073486
				-1.72946631908417 -0.773778438568115
				-1.72666752338409 -0.809341192245483
				-1.72148263454437 -0.844635128974915
				-1.71393549442291 -0.879500389099121
				-1.70405995845795 -0.913779020309448
				-1.69190096855164 -0.947315692901611
				-1.67751348018646 -0.979958295822144
				-1.6609628200531 -1.01155936717987
				-1.64232385158539 -1.04197537899017
				-1.62168097496033 -1.07106876373291
				-1.5991278886795 -1.09870755672455
				-1.57476663589478 -1.12476658821106
				-1.54870748519897 -1.14912784099579
				-1.52106869220734 -1.1716810464859
				-1.49197542667389 -1.19232380390167
				-1.46155941486359 -1.21096277236938
				-1.42995834350586 -1.22751355171204
				-1.39731562137604 -1.24190092086792
				-1.36377894878387 -1.25405991077423
				-1.32950043678284 -1.2639354467392
				-1.29463517665863 -1.27148270606995
				-1.25934112071991 -1.27666747570038
				-1.22377836704254 -1.27946627140045
				-1.17995047569275 -1.27999997138977
				0.143316864967346 -1.27999997138977
			};
		\addplot [semithick, black, forget plot]
		table {%
				0.149999976158142 -0.819999933242798
				1.21095037460327 -0.819138169288635
				1.22163116931915 -0.816573977470398
				1.23177933692932 -0.812370538711548
				1.24114501476288 -0.806631207466125
				1.24949753284454 -0.799497485160828
				1.25663113594055 -0.791144967079163
				1.26237046718597 -0.781779289245605
				1.26657390594482 -0.771631240844727
				1.26913821697235 -0.760950326919556
				1.26999998092651 -0.75
				1.26913821697235 0.760950326919556
				1.26657390594482 0.771631240844727
				1.26237046718597 0.781779289245605
				1.25663113594055 0.791144967079163
				1.24949753284454 0.799497485160828
				1.24114501476288 0.806631207466125
				1.23177933692932 0.812370538711548
				1.22163116931915 0.816573977470398
				1.21095037460327 0.819138169288635
				1.20000004768372 0.819999933242798
				0.889049530029297 0.819138169288635
				0.878368854522705 0.816573977470398
				0.868220686912537 0.812370538711548
				0.858855009078979 0.806631207466125
				0.850502490997314 0.799497485160828
				0.843368768692017 0.791144967079163
				0.837629556655884 0.781779289245605
				0.833425998687744 0.771631240844727
				0.830861806869507 0.760950326919556
				0.829999923706055 0.75
				0.829466342926025 0.126221656799316
				0.826667547225952 0.0906587839126587
				0.82148265838623 0.0553648471832275
				0.813935518264771 0.020499587059021
				0.804059982299805 -0.0137790441513062
				0.791900873184204 -0.0473155975341797
				0.77751350402832 -0.079958438873291
				0.760962724685669 -0.111559391021729
				0.742323875427246 -0.141975402832031
				0.721680998802185 -0.171068668365479
				0.699127912521362 -0.198707580566406
				0.674766540527344 -0.224766612052917
				0.648707509040833 -0.249127864837646
				0.621068716049194 -0.271681070327759
				0.591975450515747 -0.29232382774353
				0.561559438705444 -0.310962796211243
				0.529958367347717 -0.327513456344604
				0.497315645217896 -0.341900944709778
				0.463778972625732 -0.354059934616089
				0.429500341415405 -0.363935470581055
				0.394635200500488 -0.371482610702515
				0.359341144561768 -0.376667499542236
				0.323778390884399 -0.37946629524231
				0.279999971389771 -0.379999995231628
				-0.0237783193588257 -0.37946629524231
				-0.0593411922454834 -0.376667499542236
				-0.0946351289749146 -0.371482610702515
				-0.129500389099121 -0.363935470581055
				-0.163779020309448 -0.354059934616089
				-0.197315692901611 -0.341900944709778
				-0.229958415031433 -0.327513456344604
				-0.261559367179871 -0.310962796211243
				-0.291975378990173 -0.29232382774353
				-0.32106876373291 -0.271681070327759
				-0.348707556724548 -0.249127864837646
				-0.37476658821106 -0.224766612052917
				-0.399127840995789 -0.198707580566406
				-0.421681046485901 -0.171068668365479
				-0.442323803901672 -0.141975402832031
				-0.460962772369385 -0.111559391021729
				-0.477513551712036 -0.079958438873291
				-0.49190092086792 -0.0473155975341797
				-0.504060029983521 -0.0137790441513062
				-0.513935446739197 0.020499587059021
				-0.521482706069946 0.0553648471832275
				-0.526667475700378 0.0906587839126587
				-0.529466390609741 0.126221656799316
				-0.529999971389771 0.169999957084656
				-0.530861854553223 0.760950326919556
				-0.53342604637146 0.771631240844727
				-0.5376296043396 0.781779289245605
				-0.543368816375732 0.791144967079163
				-0.55050253868103 0.799497485160828
				-0.558855056762695 0.806631207466125
				-0.568220615386963 0.812370538711548
				-0.578368782997131 0.816573977470398
				-0.589049577713013 0.819138169288635
				-0.600000023841858 0.819999933242798
				-1.21095037460327 0.819138169288635
				-1.22163116931915 0.816573977470398
				-1.23177933692932 0.812370538711548
				-1.24114501476288 0.806631207466125
				-1.24949753284454 0.799497485160828
				-1.25663113594055 0.791144967079163
				-1.26237046718597 0.781779289245605
				-1.26657390594482 0.771631240844727
				-1.26913821697235 0.760950326919556
				-1.26999998092651 0.75
				-1.26913821697235 -0.760950326919556
				-1.26657390594482 -0.771631240844727
				-1.26237046718597 -0.781779289245605
				-1.25663113594055 -0.791144967079163
				-1.24949753284454 -0.799497485160828
				-1.24114501476288 -0.806631207466125
				-1.23177933692932 -0.812370538711548
				-1.22163116931915 -0.816573977470398
				-1.21095037460327 -0.819138169288635
				-1.20000004768372 -0.819999933242798
				1.21095037460327 -0.819138169288635
				1.22163116931915 -0.816573977470398
				1.23177933692932 -0.812370538711548
				1.24114501476288 -0.806631207466125
				1.24949753284454 -0.799497485160828
				1.25663113594055 -0.791144967079163
				1.26237046718597 -0.781779289245605
				1.26657390594482 -0.771631240844727
				1.26913821697235 -0.760950326919556
				1.26999998092651 -0.75
				1.26913821697235 0.760950326919556
				1.26657390594482 0.771631240844727
				1.26237046718597 0.781779289245605
				1.25663113594055 0.791144967079163
				1.24949753284454 0.799497485160828
				1.24114501476288 0.806631207466125
				1.23177933692932 0.812370538711548
				1.22163116931915 0.816573977470398
				1.21095037460327 0.819138169288635
				1.20000004768372 0.819999933242798
				0.889049530029297 0.819138169288635
				0.878368854522705 0.816573977470398
				0.868220686912537 0.812370538711548
				0.858855009078979 0.806631207466125
				0.850502490997314 0.799497485160828
				0.843368768692017 0.791144967079163
				0.837629556655884 0.781779289245605
				0.833425998687744 0.771631240844727
				0.830861806869507 0.760950326919556
				0.829999923706055 0.75
				0.829466342926025 0.126221656799316
				0.826667547225952 0.0906587839126587
				0.82148265838623 0.0553648471832275
				0.813935518264771 0.020499587059021
				0.804059982299805 -0.0137790441513062
				0.791900873184204 -0.0473155975341797
				0.77751350402832 -0.079958438873291
				0.760962724685669 -0.111559391021729
				0.742323875427246 -0.141975402832031
				0.721680998802185 -0.171068668365479
				0.699127912521362 -0.198707580566406
				0.674766540527344 -0.224766612052917
				0.648707509040833 -0.249127864837646
				0.621068716049194 -0.271681070327759
				0.591975450515747 -0.29232382774353
				0.561559438705444 -0.310962796211243
				0.529958367347717 -0.327513456344604
				0.497315645217896 -0.341900944709778
				0.463778972625732 -0.354059934616089
				0.429500341415405 -0.363935470581055
				0.394635200500488 -0.371482610702515
				0.359341144561768 -0.376667499542236
				0.323778390884399 -0.37946629524231
				0.279999971389771 -0.379999995231628
				-0.0237783193588257 -0.37946629524231
				-0.0593411922454834 -0.376667499542236
				-0.0946351289749146 -0.371482610702515
				-0.129500389099121 -0.363935470581055
				-0.163779020309448 -0.354059934616089
				-0.197315692901611 -0.341900944709778
				-0.229958415031433 -0.327513456344604
				-0.261559367179871 -0.310962796211243
				-0.291975378990173 -0.29232382774353
				-0.32106876373291 -0.271681070327759
				-0.348707556724548 -0.249127864837646
				-0.37476658821106 -0.224766612052917
				-0.399127840995789 -0.198707580566406
				-0.421681046485901 -0.171068668365479
				-0.442323803901672 -0.141975402832031
				-0.460962772369385 -0.111559391021729
				-0.477513551712036 -0.079958438873291
				-0.49190092086792 -0.0473155975341797
				-0.504060029983521 -0.0137790441513062
				-0.513935446739197 0.020499587059021
				-0.521482706069946 0.0553648471832275
				-0.526667475700378 0.0906587839126587
				-0.529466390609741 0.126221656799316
				-0.529999971389771 0.169999957084656
				-0.530861854553223 0.760950326919556
				-0.53342604637146 0.771631240844727
				-0.5376296043396 0.781779289245605
				-0.543368816375732 0.791144967079163
				-0.55050253868103 0.799497485160828
				-0.558855056762695 0.806631207466125
				-0.568220615386963 0.812370538711548
				-0.578368782997131 0.816573977470398
				-0.589049577713013 0.819138169288635
				-0.600000023841858 0.819999933242798
				-1.21095037460327 0.819138169288635
				-1.22163116931915 0.816573977470398
				-1.23177933692932 0.812370538711548
				-1.24114501476288 0.806631207466125
				-1.24949753284454 0.799497485160828
				-1.25663113594055 0.791144967079163
				-1.26237046718597 0.781779289245605
				-1.26657390594482 0.771631240844727
				-1.26913821697235 0.760950326919556
				-1.26999998092651 0.75
				-1.26913821697235 -0.760950326919556
				-1.26657390594482 -0.771631240844727
				-1.26237046718597 -0.781779289245605
				-1.25663113594055 -0.791144967079163
				-1.24949753284454 -0.799497485160828
				-1.24114501476288 -0.806631207466125
				-1.23177933692932 -0.812370538711548
				-1.22163116931915 -0.816573977470398
				-1.21095037460327 -0.819138169288635
				-1.20000004768372 -0.819999933242798
				0.143316864967346 -0.819999933242798
			};
		\addplot [semithick, red, dashed]
		table {%
				0.149999976158142 -1.04999995231628
				0.282364845275879 -1.0490962266922
				0.490423440933228 -1.04574012756348
				0.683572053909302 -1.04208624362946
				0.884199619293213 -1.03977417945862
				0.945521593093872 -1.03635942935944
				0.992515087127686 -1.03186285495758
				1.04015898704529 -1.02505564689636
				1.0721207857132 -1.01897156238556
				1.10412573814392 -1.01146149635315
				1.13605296611786 -1.00240743160248
				1.16773962974548 -0.991705894470215
				1.1990602016449 -0.979285359382629
				1.22986960411072 -0.965099573135376
				1.26000416278839 -0.949127674102783
				1.28933846950531 -0.931370973587036
				1.31774139404297 -0.91185474395752
				1.3450835943222 -0.890632390975952
				1.3712751865387 -0.86776328086853
				1.39622449874878 -0.843320965766907
				1.41986048221588 -0.817404985427856
				1.44214844703674 -0.790104746818542
				1.46305239200592 -0.761515617370605
				1.48256254196167 -0.731748580932617
				1.50069642066956 -0.700896382331848
				1.51746308803558 -0.669050693511963
				1.53289413452148 -0.6363126039505
				1.54703891277313 -0.602758049964905
				1.55993723869324 -0.568457841873169
				1.57708728313446 -0.515760898590088
				1.59178304672241 -0.46171772480011
				1.60424149036407 -0.406511187553406
				1.6146901845932 -0.350285291671753
				1.62333202362061 -0.293160796165466
				1.63039112091064 -0.235222101211548
				1.63767147064209 -0.15682053565979
				1.64284574985504 -0.0771738290786743
				1.64605402946472 0.0036848783493042
				1.64722204208374 0.0857124328613281
				1.64607203006744 0.1688392162323
				1.64340698719025 0.231825232505798
				1.6389068365097 0.295258164405823
				1.63223266601562 0.359017372131348
				1.62300300598145 0.42293381690979
				1.61520743370056 0.465512275695801
				1.605952501297 0.507965326309204
				1.59509801864624 0.550185203552246
				1.58251941204071 0.592032551765442
				1.56810212135315 0.633350133895874
				1.5517121553421 0.673936128616333
				1.53329133987427 0.713591337203979
				1.51279830932617 0.752112865447998
				1.49021899700165 0.78929328918457
				1.46557986736298 0.82490599155426
				1.43894255161285 0.858753442764282
				1.41039729118347 0.890650987625122
				1.38006615638733 0.920425891876221
				1.34808778762817 0.947933673858643
				1.31462013721466 0.973044633865356
				1.2798376083374 0.995648622512817
				1.24391841888428 1.01566016674042
				1.2070506811142 1.03301024436951
				1.16943001747131 1.04764902591705
				1.13125026226044 1.05954623222351
				1.09271359443665 1.06868696212769
				1.05402183532715 1.07507359981537
				1.01536810398102 1.07872569561005
				0.976954936981201 1.07967698574066
				0.938971757888794 1.07797622680664
				0.901589632034302 1.07368516921997
				0.864984512329102 1.06687879562378
				0.829320907592773 1.05764293670654
				0.794752240180969 1.04607403278351
				0.761435270309448 1.03228211402893
				0.729514360427856 1.01638650894165
				0.699123382568359 0.998514890670776
				0.670378923416138 0.978799819946289
				0.643391728401184 0.957383155822754
				0.618257164955139 0.934409618377686
				0.595046877861023 0.910032749176025
				0.573811411857605 0.884418725967407
				0.554558753967285 0.857745409011841
				0.537304639816284 0.830167531967163
				0.522000432014465 0.801861643791199
				0.508596420288086 0.772982835769653
				0.497004270553589 0.743685364723206
				0.487106084823608 0.714109420776367
				0.478801727294922 0.684368848800659
				0.471966981887817 0.654569864273071
				0.46420431137085 0.609941840171814
				0.459080457687378 0.565577983856201
				0.456161022186279 0.521594047546387
				0.455533981323242 0.478346705436707
				0.457194209098816 0.422325134277344
				0.464644074440002 0.264233946800232
				0.463862419128418 0.2144775390625
				0.461294889450073 0.178081393241882
				0.456646800041199 0.142499923706055
				0.449602127075195 0.107815265655518
				0.443436741828918 0.085270881652832
				0.43600594997406 0.0632607936859131
				0.427263975143433 0.0418517589569092
				0.417208313941956 0.021108865737915
				0.405825614929199 0.00107753276824951
				0.393131375312805 -0.0181833505630493
				0.379164695739746 -0.0365918874740601
				0.363958120346069 -0.0540904998779297
				0.347553014755249 -0.0706202983856201
				0.330000400543213 -0.0861120223999023
				0.311350584030151 -0.100505471229553
				0.291653394699097 -0.113744258880615
				0.270970106124878 -0.125760912895203
				0.24936306476593 -0.136495471000671
				0.226894855499268 -0.145891904830933
				0.203640341758728 -0.153886556625366
				0.179674983024597 -0.160424113273621
				0.155078291893005 -0.165453195571899
				0.129940986633301 -0.16891884803772
				0.104352593421936 -0.170776128768921
				0.0784087181091309 -0.170984506607056
				0.0522153377532959 -0.169503688812256
				0.0258771181106567 -0.166304349899292
				-0.000491738319396973 -0.16136372089386
				-0.0267652273178101 -0.154664635658264
				-0.0528093576431274 -0.146204233169556
				-0.0784988403320312 -0.136006951332092
				-0.103706002235413 -0.124099731445312
				-0.128310441970825 -0.110540747642517
				-0.152203679084778 -0.0953958034515381
				-0.175285458564758 -0.0787315368652344
				-0.19747006893158 -0.0606398582458496
				-0.218682527542114 -0.0412138700485229
				-0.238850593566895 -0.0205405950546265
				-0.257909178733826 0.00127267837524414
				-0.275827169418335 0.0241093635559082
				-0.292569756507874 0.0478705167770386
				-0.308125376701355 0.0724349021911621
				-0.322523355484009 0.0976787805557251
				-0.341963648796082 0.136610269546509
				-0.358887791633606 0.1765296459198
				-0.373353719711304 0.217171669006348
				-0.385362982749939 0.258330821990967
				-0.394911050796509 0.299821496009827
				-0.402015805244446 0.341473937034607
				-0.406759977340698 0.383145689964294
				-0.409403800964355 0.424742102622986
				-0.410201907157898 0.466189384460449
				-0.40917432308197 0.521188378334045
				-0.405668497085571 0.644548416137695
				-0.406877875328064 0.685690760612488
				-0.410488843917847 0.726691961288452
				-0.414511322975159 0.753832340240479
				-0.419964075088501 0.780715584754944
				-0.426958680152893 0.807239294052124
				-0.435552358627319 0.833307981491089
				-0.445772767066956 0.858816027641296
				-0.45763373374939 0.883667945861816
				-0.471125364303589 0.907765626907349
				-0.48621940612793 0.931006908416748
				-0.502882957458496 0.953304529190063
				-0.521069526672363 0.974566221237183
				-0.540721416473389 0.994701266288757
				-0.56177806854248 1.01363635063171
				-0.584163904190063 1.03129422664642
				-0.607799887657166 1.04759788513184
				-0.632607936859131 1.06249165534973
				-0.658497333526611 1.0759152173996
				-0.685377597808838 1.08780860900879
				-0.713162183761597 1.09812784194946
				-0.741752624511719 1.10681867599487
				-0.771049737930298 1.11383128166199
				-0.800957679748535 1.11912286281586
				-0.831366300582886 1.12264060974121
				-0.862164735794067 1.12434077262878
				-0.89324152469635 1.12418293952942
				-0.924470186233521 1.12212240695953
				-0.955723285675049 1.11812651157379
				-0.986869096755981 1.11216473579407
				-1.01776266098022 1.10421419143677
				-1.0482622385025 1.09426844120026
				-1.07822775840759 1.0823312997818
				-1.10751116275787 1.06842088699341
				-1.13596880435944 1.05256795883179
				-1.16346502304077 1.03481733798981
				-1.18986463546753 1.01522982120514
				-1.21504390239716 0.993875980377197
				-1.23889994621277 0.970840454101562
				-1.26134216785431 0.946224689483643
				-1.28228569030762 0.920130133628845
				-1.30167543888092 0.892667055130005
				-1.31947636604309 0.863955497741699
				-1.33565616607666 0.834107637405396
				-1.35021603107452 0.803235292434692
				-1.36317956447601 0.771453619003296
				-1.37456285953522 0.738864183425903
				-1.38441610336304 0.705564618110657
				-1.39279735088348 0.671648502349854
				-1.39975547790527 0.637195587158203
				-1.40536308288574 0.602281332015991
				-1.41139662265778 0.549192190170288
				-1.41480886936188 0.495407462120056
				-1.41590762138367 0.44108784198761
				-1.4150482416153 0.386357188224792
				-1.41162657737732 0.312883138656616
				-1.39589107036591 0.052207350730896
				-1.3947411775589 -0.0230153799057007
				-1.39607787132263 -0.098374605178833
				-1.39966213703156 -0.173809289932251
				-1.4087301492691 -0.30608856678009
				-1.41296327114105 -0.381985425949097
				-1.41424584388733 -0.439039349555969
				-1.41374802589417 -0.477065205574036
				-1.41190147399902 -0.514995574951172
				-1.40856087207794 -0.552747964859009
				-1.40368092060089 -0.590264558792114
				-1.39719212055206 -0.627455234527588
				-1.38902342319489 -0.664203286170959
				-1.37916588783264 -0.700413942337036
				-1.36758840084076 -0.735974550247192
				-1.35425269603729 -0.77074658870697
				-1.33914804458618 -0.804604053497314
				-1.32226431369781 -0.837408542633057
				-1.30360436439514 -0.869001746177673
				-1.28320562839508 -0.89924693107605
				-1.26112949848175 -0.928030490875244
				-1.2374621629715 -0.955250978469849
				-1.21230435371399 -0.980827569961548
				-1.18577039241791 -1.00471866130829
				-1.15798497200012 -1.02690804004669
				-1.12906980514526 -1.04739022254944
				-1.09914529323578 -1.06619644165039
				-1.06832718849182 -1.08337187767029
				-1.03671967983246 -1.09896111488342
				-1.00441837310791 -1.11303675174713
				-0.971509695053101 -1.12566912174225
				-0.938068628311157 -1.13692080974579
				-0.904159545898438 -1.14687347412109
				-0.85254430770874 -1.15954434871674
				-0.800161838531494 -1.16972017288208
				-0.747132539749146 -1.17759788036346
				-0.69354510307312 -1.18331575393677
				-0.639467358589172 -1.1871223449707
				-0.584949493408203 -1.18910670280457
				-0.530009031295776 -1.18944847583771
				-0.474663734436035 -1.18821227550507
				-0.400241136550903 -1.18425238132477
				-0.325121879577637 -1.17772424221039
				-0.249313592910767 -1.16876542568207
				-0.172827363014221 -1.15758800506592
				-0.0763169527053833 -1.14101696014404
				0.0604375600814819 -1.11496555805206
				0.119682431221008 -1.10365998744965
			};
		%\addlegendentry{Ground Truth}
		\addplot [semithick, color0]
		table {%
				0.149999976158142 -1.04999995231628
				0.260098695755005 -1.04915392398834
				0.539182901382446 -1.04320299625397
				0.808214902877808 -1.03549647331238
				0.882252931594849 -1.03136157989502
				0.927961111068726 -1.02738046646118
				0.974517345428467 -1.0216578245163
				1.02171921730042 -1.01370000839233
				1.06930160522461 -1.00300443172455
				1.10106074810028 -0.994155406951904
				1.13271570205688 -0.983810186386108
				1.16413879394531 -0.971895694732666
				1.19518983364105 -0.958337306976318
				1.22573471069336 -0.943096160888672
				1.25562453269958 -0.926146984100342
				1.28472578525543 -0.907479643821716
				1.31290602684021 -0.887109637260437
				1.34003686904907 -0.865079164505005
				1.36599552631378 -0.841431617736816
				1.3907059431076 -0.816241860389709
				1.41409289836884 -0.789601802825928
				1.4361172914505 -0.761600732803345
				1.45674228668213 -0.732333779335022
				1.47595977783203 -0.701912403106689
				1.49379181861877 -0.670432806015015
				1.51025354862213 -0.637985587120056
				1.52538537979126 -0.604667901992798
				1.53924250602722 -0.570554256439209
				1.55186855792999 -0.535712242126465
				1.56863737106323 -0.482225060462952
				1.58295512199402 -0.427430987358093
				1.59502398967743 -0.371500492095947
				1.60504007339478 -0.314575433731079
				1.6132310628891 -0.256770372390747
				1.61982560157776 -0.19817042350769
				1.62650680541992 -0.118880271911621
				1.6310567855835 -0.038316011428833
				1.63355004787445 0.0434830188751221
				1.63384854793549 0.12646496295929
				1.6324360370636 0.189409255981445
				1.62940466403961 0.252897024154663
				1.62445557117462 0.316830992698669
				1.61723411083221 0.381070971488953
				1.60734510421753 0.445444226264954
				1.59906303882599 0.488313436508179
				1.58926558494568 0.531030535697937
				1.57784259319305 0.573489189147949
				1.56466794013977 0.615555882453918
				1.54960334300995 0.657042026519775
				1.53255951404572 0.69775652885437
				1.51347875595093 0.737510204315186
				1.4923267364502 0.776093006134033
				1.46910309791565 0.813283920288086
				1.44384336471558 0.84889018535614
				1.41662073135376 0.882718324661255
				1.38754367828369 0.91458523273468
				1.35673594474792 0.94434118270874
				1.32434511184692 0.971848368644714
				1.29053604602814 0.996987819671631
				1.25546979904175 1.01967060565948
				1.21933436393738 1.03981482982635
				1.18233215808868 1.05735313892365
				1.14465200901031 1.07224535942078
				1.10649538040161 1.08446335792542
				1.0680673122406 1.0939964056015
				1.02954936027527 1.10085451602936
				0.991121053695679 1.10506272315979
				0.952994585037231 1.10665655136108
				0.915368914604187 1.10568845272064
				0.87841272354126 1.1022264957428
				0.84235417842865 1.09635603427887
				0.807344913482666 1.08817148208618
				0.773529052734375 1.07777881622314
				0.741078615188599 1.06530404090881
				0.710144996643066 1.0508828163147
				0.680853366851807 1.03465819358826
				0.653296232223511 1.01677429676056
				0.627566337585449 0.997383832931519
				0.603718638420105 0.976623773574829
				0.581794023513794 0.954637289047241
				0.561819553375244 0.931557893753052
				0.543820381164551 0.907525539398193
				0.527810573577881 0.882682085037231
				0.513796806335449 0.857160091400146
				0.501708984375 0.831101655960083
				0.491485595703125 0.804638862609863
				0.482996940612793 0.777888059616089
				0.476062655448914 0.750958561897278
				0.468199253082275 0.710410356521606
				0.462864637374878 0.669876337051392
				0.45953369140625 0.629506349563599
				0.457571983337402 0.576090812683105
				0.457744359970093 0.523259401321411
				0.460002422332764 0.458005309104919
				0.46474552154541 0.341894507408142
				0.464557409286499 0.290481209754944
				0.462862849235535 0.251968860626221
				0.45961856842041 0.213504552841187
				0.454330205917358 0.175252079963684
				0.446547627449036 0.137428760528564
				0.439748167991638 0.112617135047913
				0.431547522544861 0.088257908821106
				0.42190957069397 0.0644652843475342
				0.410798311233521 0.0413720607757568
				0.398226022720337 0.0190836191177368
				0.384225487709045 -0.00229227542877197
				0.368844747543335 -0.022651195526123
				0.352147102355957 -0.0418988466262817
				0.33420729637146 -0.0599470138549805
				0.315108895301819 -0.0767123699188232
				0.294939875602722 -0.0921217203140259
				0.273794412612915 -0.106107831001282
				0.251771211624146 -0.11860990524292
				0.228967666625977 -0.129576563835144
				0.205487132072449 -0.138962268829346
				0.181435227394104 -0.146727561950684
				0.156913995742798 -0.152841687202454
				0.132031083106995 -0.157279133796692
				0.106894373893738 -0.16002094745636
				0.0816059112548828 -0.161055445671082
				0.0562736988067627 -0.160375952720642
				0.0310032367706299 -0.157982110977173
				0.00590264797210693 -0.15388011932373
				-0.0189173221588135 -0.14808201789856
				-0.0433521270751953 -0.14060640335083
				-0.067293643951416 -0.131478548049927
				-0.0906312465667725 -0.120730400085449
				-0.113266348838806 -0.108402252197266
				-0.135095477104187 -0.094546914100647
				-0.156030893325806 -0.0792227983474731
				-0.17597496509552 -0.0624895095825195
				-0.19485604763031 -0.0444290637969971
				-0.212597727775574 -0.0251154899597168
				-0.229145526885986 -0.00462555885314941
				-0.244462251663208 0.0169486999511719
				-0.258504509925842 0.0395194292068481
				-0.271259069442749 0.062991738319397
				-0.282738208770752 0.0872671604156494
				-0.292935729026794 0.112261295318604
				-0.301881074905396 0.137891411781311
				-0.313059091567993 0.177351474761963
				-0.321699738502502 0.217844367027283
				-0.328053951263428 0.2591792345047
				-0.332426190376282 0.301243185997009
				-0.335852742195129 0.358329176902771
				-0.337771892547607 0.431187033653259
				-0.340332388877869 0.52081561088562
				-0.343398809432983 0.566437959671021
				-0.348630428314209 0.612399816513062
				-0.353658199310303 0.643095970153809
				-0.360102772712708 0.673722505569458
				-0.368117809295654 0.704156637191772
				-0.377829551696777 0.734261512756348
				-0.389314532279968 0.763907670974731
				-0.402617692947388 0.79295015335083
				-0.417763471603394 0.821242809295654
				-0.434739589691162 0.848632335662842
				-0.453518748283386 0.874991059303284
				-0.474052429199219 0.900205612182617
				-0.496273279190063 0.924145460128784
				-0.520102024078369 0.94671094417572
				-0.545453071594238 0.967813014984131
				-0.572233200073242 0.987373352050781
				-0.600341796875 1.00530958175659
				-0.629673719406128 1.02155363559723
				-0.660120248794556 1.03604245185852
				-0.691569566726685 1.04871392250061
				-0.723905324935913 1.05951118469238
				-0.757011413574219 1.06838631629944
				-0.790768384933472 1.07529044151306
				-0.825052976608276 1.08019256591797
				-0.859742403030396 1.08306300640106
				-0.894714951515198 1.0838862657547
				-0.929846286773682 1.08265650272369
				-0.965008497238159 1.07936596870422
				-1.00006783008575 1.07401072978973
				-1.03489422798157 1.06660044193268
				-1.06935632228851 1.05714643001556
				-1.10331583023071 1.04566562175751
				-1.13664376735687 1.03218650817871
				-1.16921257972717 1.01674377918243
				-1.20089590549469 0.999385833740234
				-1.23158359527588 0.980175375938416
				-1.26117014884949 0.959180593490601
				-1.28957045078278 0.936493396759033
				-1.31671118736267 0.912205338478088
				-1.34254419803619 0.886420965194702
				-1.36702716350555 0.859239101409912
				-1.39014327526093 0.830767273902893
				-1.41189575195312 0.801113128662109
				-1.43227362632751 0.770366430282593
				-1.45128273963928 0.738617062568665
				-1.46895968914032 0.705962777137756
				-1.48533010482788 0.672487258911133
				-1.50041377544403 0.638262271881104
				-1.5143004655838 0.603373050689697
				-1.53298342227936 0.549915194511414
				-1.54925847053528 0.495252847671509
				-1.56327879428864 0.439522981643677
				-1.57519197463989 0.38284707069397
				-1.58511757850647 0.325324535369873
				-1.5932149887085 0.267046213150024
				-1.59966230392456 0.208082556724548
				-1.6060026884079 0.128495097160339
				-1.61003541946411 0.0478906631469727
				-1.61180770397186 -0.0336507558822632
				-1.61153304576874 -0.0953747034072876
				-1.60968887805939 -0.157530426979065
				-1.60603737831116 -0.220041036605835
				-1.60028326511383 -0.282773375511169
				-1.59212839603424 -0.345547318458557
				-1.58123874664307 -0.408137798309326
				-1.57227289676666 -0.44959545135498
				-1.56182169914246 -0.490713715553284
				-1.5497841835022 -0.531362414360046
				-1.53608000278473 -0.571412086486816
				-1.52063632011414 -0.610716700553894
				-1.50341594219208 -0.649126410484314
				-1.48440623283386 -0.686500310897827
				-1.46361744403839 -0.722700357437134
				-1.44108593463898 -0.757606029510498
				-1.41687321662903 -0.791105508804321
				-1.39105439186096 -0.823107838630676
				-1.3637181520462 -0.853537917137146
				-1.33497297763824 -0.882349252700806
				-1.30492579936981 -0.909509420394897
				-1.27369284629822 -0.935014009475708
				-1.24138355255127 -0.958864808082581
				-1.20810878276825 -0.981087565422058
				-1.17397737503052 -1.00172817707062
				-1.13908493518829 -1.02082800865173
				-1.10352313518524 -1.03844678401947
				-1.06737780570984 -1.05465197563171
				-1.03072214126587 -1.06950211524963
				-0.974928855895996 -1.08939576148987
				-0.918335914611816 -1.10663878917694
				-0.861102700233459 -1.12145793437958
				-0.803360462188721 -1.13407599925995
				-0.745218515396118 -1.14470756053925
				-0.686763763427734 -1.15357637405396
				-0.608450889587402 -1.16298711299896
				-0.529830098152161 -1.17010509967804
				-0.450989723205566 -1.17538440227509
				-0.352218151092529 -1.18001878261566
				-0.23345935344696 -1.1834819316864
				-0.0946369171142578 -1.18554902076721
				0.02475905418396 -1.18569242954254
				0.0846474170684814 -1.18500900268555
			};
		%\addlegendentry{Alpaca}
	\end{axis}

\end{tikzpicture}

%% file: car_lap_rmse.tex
%!TEX root = ../neurips_2020.tex

% This file was created by tikzplotlib v0.8.5.
\begin{tikzpicture}

	\definecolor{color0}{rgb}{1,0.498039215686275,0.0549019607843137}

	\begin{axis}[
			tick align=outside,
			tick pos=left,
			x grid style={white!69.01960784313725!black},
			xmin=0.5, xmax=3.5,
			xtick style={color=black},
			xtick={1,2,3},
			xticklabels={MMPC,MMPC-GP,Alpaca},
			y grid style={white!69.01960784313725!black},
			ymin=-0.000150185917127635, ymax=0.0420578934246638,
			ytick style={color=black},
			scale={0.4},
			x = {3.5cm},
			ylabel = {RMSE}
		]
		\addplot [black]
		table {%
				0.85 0.00247382274522332
				1.15 0.00247382274522332
				1.15 0.0054576954563483
				0.85 0.0054576954563483
				0.85 0.00247382274522332
			};
		\addplot [black]
		table {%
				1 0.00247382274522332
				1 0.00176836314386289
			};
		\addplot [black]
		table {%
				1 0.0054576954563483
				1 0.00988470177254751
			};
		\addplot [black]
		table {%
				0.925 0.00176836314386289
				1.075 0.00176836314386289
			};
		\addplot [black]
		table {%
				0.925 0.00988470177254751
				1.075 0.00988470177254751
			};
		\addplot [black, mark=*, mark size=3, mark options={solid,fill opacity=0}, only marks]
		table {%
				1 0.0103334566632371
				1 0.0123345345342212
				1 0.0101323444681571
				1 0.0115679597287273
				1 0.0146132880577516
				1 0.0101570569648993
				1 0.0102454292349551
				1 0.0113563332024136
				1 0.0109415984606954
				1 0.0111681942669303
				1 0.0101473074056274
				1 0.0100551298514341
				1 0.0105548439246607
				1 0.0102003142030902
				1 0.0121905436739951
				1 0.0100328969545953
				1 0.0103476837060049
				1 0.0101432361592867
				1 0.0109574213116016
				1 0.0099453044227877
				1 0.0103270246547517
				1 0.0105910423810621
				1 0.0108966781694338
				1 0.0106058610911885
				1 0.0110157401241214
				1 0.0108025057107328
				1 0.0105192317544843
				1 0.0101993089068085
				1 0.0108687444481052
			};
		\addplot [black]
		table {%
				1.85 0.00619293857601991
				2.15 0.00619293857601991
				2.15 0.0111450682810839
				1.85 0.0111450682810839
				1.85 0.00619293857601991
			};
		\addplot [black]
		table {%
				2 0.00619293857601991
				2 0.00183839478483021
			};
		\addplot [black]
		table {%
				2 0.0111450682810839
				2 0.0181957483927381
			};
		\addplot [black]
		table {%
				1.925 0.00183839478483021
				2.075 0.00183839478483021
			};
		\addplot [black]
		table {%
				1.925 0.0181957483927381
				2.075 0.0181957483927381
			};
		\addplot [black, mark=*, mark size=3, mark options={solid,fill opacity=0}, only marks]
		table {%
				2 0.0198369288461108
				2 0.0236166786365341
				2 0.0265677078340639
				2 0.0257705039857654
				2 0.0283548426526865
				2 0.0285546331258838
				2 0.0244601165402464
				2 0.0264988142443063
				2 0.0262014605771542
				2 0.0275525083681853
				2 0.0279768962384205
				2 0.0282819476016621
				2 0.0250601821284982
				2 0.0277772468355103
				2 0.0276116382202846
				2 0.0242968219905307
				2 0.0240032654347368
				2 0.0265566966295202
			};
		\addplot [black]
		table {%
				2.85 0.0266726754781975
				3.15 0.0266726754781975
				3.15 0.0336298503352295
				2.85 0.0336298503352295
				2.85 0.0266726754781975
			};
		\addplot [black]
		table {%
				3 0.0266726754781975
				3 0.0162535630381607
			};
		\addplot [black]
		table {%
				3 0.0336298503352295
				3 0.0401393443636733
			};
		\addplot [black]
		table {%
				2.925 0.0162535630381607
				3.075 0.0162535630381607
			};
		\addplot [black]
		table {%
				2.925 0.0401393443636733
				3.075 0.0401393443636733
			};
		\addplot [black, mark=*, mark size=3, mark options={solid,fill opacity=0}, only marks]
		table {%
				3 0.0066150413292653
				3 0.0162149015491467
				3 0.0160873636668786
				3 0.0105299653693565
				3 0.0138355331782671
				3 0.00244674311294618
				3 0.0135336773603272
				3 0.0107680765887406
				3 0.00885343658756828
				3 0.014494596221801
				3 0.0141472918591895
				3 0.0131065691879379
				3 0.0088090364119504
				3 0.0132214946252801
				3 0.00825331943963789
				3 0.0160608294402864
				3 0.0150097369170916
				3 0.0124817394774947
				3 0.00234896193260177
				3 0.0154436493264264
				3 0.0151017427356295
				3 0.0137205451077963
				3 0.00682573051568036
				3 0.0139961362043007
				3 0.0148532465768258
				3 0.00856511201886878
				3 0.00927788745857361
			};
		\addplot [color0]
		table {%
				0.85 0.00376050235673044
				1.15 0.00376050235673044
			};
		\addplot [color0]
		table {%
				1.85 0.00842342342456724
				2.15 0.00842342342456724
			};
		\addplot [color0]
		table {%
				2.85 0.0304117958201625
				3.15 0.0304117958201625
			};
	\end{axis}

\end{tikzpicture}